\documentclass{aa}

\usepackage{graphicx}
\usepackage{txfonts}
\usepackage{amsmath}	
\usepackage{newtxtext,newtxmath}
\usepackage{graphicx}
\usepackage{dcolumn}
\usepackage{bm}
\usepackage{mathtools}
\usepackage{float}
\usepackage{lineno}
\usepackage{caption}
\usepackage{booktabs}
\usepackage{array}
\usepackage{geometry}
\usepackage{subcaption}
\usepackage{physics}
\usepackage[dvipsnames]{xcolor}
\usepackage{natbib}
\usepackage{hyperref}
\hypersetup{colorlinks = true,
	allcolors = blue,}
\bibpunct{(}{)}{;}{a}{}{,}

\begin{document} 
	
	\title{Modeling of resistive relativistic astrophysical jets}
	
	\subtitle{Semianalytic results following a paraxial formalism}
	
	\author{Argyrios Loules
		\and
		Nektarios Vlahakis}
	
	\institute{Section of Astrophysics, Astronomy and Mechanics, Department of Physics, National and Kapodistrian University of Athens, University Campus, Zografos, GR-15784\\
		\email{arloules@phys.uoa.gr, vlahakis@phys.uoa.gr\\
		}
	}
	
	\date{}

	
	\abstract{Relativistic jets of magnetized plasma are a common high-energy astrophysical phenomenon, observed across a wide range of spatial and energy scales. In the past, semianalytic meridionally self-similar models have proven highly successful in deciphering the intricate mechanisms that determine their acceleration, collimation, and morphological characteristics. In this work, we present a modification of this formalism based on the angular expansion of the equations of general-relativistic resistive magnetohydrodynamics in the vicinity of the jet axis for the description of resistive relativistic spine jets. Our paraxial formalism allows for the inclusion of resistivity and of a realistic, variable adiabatic index equation of state in the mathematical formulation. The electric potential gradient along poloidal magnetic field lines, caused by a gradient in the rotational angular velocity of the field lines, was identified as the mechanism behind the emergence of Ohmic dissipation in resistive jets. The semianalytic solutions that we present demonstrate that Ohmic dissipation is significant only over localized dissipation regions in resistive jets. Over the extent of these regions, Ohmic dissipation weakens the thermal acceleration mechanism and can even lead to the deceleration of these outflows. Additionally, the resistive jets display enhanced collimation and a strengthening of their toroidal magnetic fields over the dissipation regions, resulting in smaller asymptotic opening angles and a more helical magnetic field structure compared to their nonresistive counterparts.}
	
	
	\maketitle
	
	
	\section{Introduction}
	
	Relativistic astrophysical jets are collimated outflows of magnetized plasma, observed across a wide spectrum of energy and spatial scales. They are ubiquitous signatures of accreting astrophysical environments, typically launched from regions where accretion of matter onto compact objects takes place. Active galactic nuclei (AGN) and X-ray binaries (XRBs) are notable examples of astrophysical environments that launch strongly relativistic jets with Lorentz factors on the order of a few tens or mildly relativistic outflows, respectively \citep{blandford2019,kylafis2012}. Additionally, ultra-relativistic jets with Lorentz factors in the hundreds are launched during short-lived cataclysmic events, such as binary neutron star mergers \citep{rezzolla2011, mpisketzis2024} or the core-collapse supernovae of massive stars \citep{pais2023}. These ultra-relativistic plasma outflows are the sources of the very high energy nonthermal emission pulses associated with these extreme astrophysical phenomena, known as gamma-ray bursts (GRBs) \citep{magic2019}. 
	
	Relativistic jets are typically launched from the vicinity of spinning black holes, and are thought to be evidence of the Blandford-Znajek mechanism in action \citep{blandford1977}. The Blandford-Znajek mechanism is a type of Penrose process of energy extraction from black holes. Matter, in the form of a thin disk or thick torus, is accreted onto the spinning black hole and drags its magnetic field with it. The accumulated magnetic field, which threads the black hole event horizon, is twisted by the rotation of spacetime near the spinning black hole, which rearranges the field into a helical configuration, which accelerates the plasma in the vicinity of the black hole's polar regions \citep{tchekhovskoy2011}. Moreover, the rotation of the magnetic field lines threading the black hole generates an electric field and consequently an outward propagating Poynting flux.
	
	Analytical modeling of astrophysical jets has traditionally been based on the assumptions of time-independence, azimuthal symmetry, and meridional self-similarity. Under these assumptions, the magnetic flux function follows a dependence $\sim\sin^{2}{\theta}$ on the polar angle $\theta$. Its dependence on the radial distance $r$ can either be expressed as a closed-form function of $r$ or be self-consistently determined by the solution of the governing equations of magnetohydrodynamics (MHD) (see \cite{tsinganos1992} and \cite{vlahakis1999} for nonrelativistic applications of this formalism). The meridionally self-similar formalism was later generalized to describe relativistic jets in the spacetime of a nonrotating Schwarzschild black hole by \cite{meliani2006}, who presented self-similar solutions of thermally accelerated and magnetically or thermally collimated relativistic outflows. This formalism has subsequently been successfully applied for the description of relativistic jets launched by rotating Kerr black holes \citep{globus2014, chantry2022}.
	
	The launching and propagation of both Newtonian and relativistic plasma outflows has been the subject of extensive and decade-spanning analytical and numerical investigation, with most studies performed in the context of ideal MHD. The modeling of resistive, Newtonian or relativistic astrophysical outflows, which is more relevant to this particular work, is sparser. \cite{fendt2002} performed resistive MHD simulations of protostellar jets launched from the turbulent accretion disks of young stellar objects (YSOs), in which the source of resistivity is the disk turbulence which is transported into the protostellar jet. \cite{cemeljic2008}, utilizing the radially self-similar solutions of disk-winds derived by \cite{vlahakis2000}, performed resistive MHD simulations of Blandford-Payne type outflows \citep{blandford1982} and studied the impact of physical electrical resistivity on the geometry of the characteristic poloidal surfaces and on the energy integral of these classical MHD outflows. An important result of this work was the identification of a regime in which resistivity is able to cause significant dissipation of electromagnetic energy at large values of the magnetic Reynolds number (i.e. without the emergence of magnetic field diffusion). In a recent work, \cite{mattia2023} performed resistive MHD simulations of relativistic jets and found that resistivity suppresses the formation of turbulence and plasmoids in their simulated relativistic flows, also causing the heating of the jet plasma through the process of Ohmic dissipation. 
	
	There is strong physical motivation for the modeling of resistive astrophysical plasma flows and outflows. Non-zero resistivity is crucial for the dissipation of magnetic energy through magnetic reconnection, a dissipative process which can efficiently accelerate particles and fuel high energy nonthermal emission in accretion flows and jets \citep{hoshino2012, rodriguez-ramirez2019, medina-torrejon2023}. Specifically, nonzero resistivity is essential for the formation of current sheets and the generation of plasmoids in the inner regions of hot accretion flows, which are thought to be the mechanism behind the high-energy emission observed during flaring events from Sgr A* \citep{nathanail2022}. More directly related to this work, dissipation of magnetic energy through magnetic reconnection can also occur in relativistic jets, far from their launching sites, where instabilities, in particular the current-driven kink instability, can foster the generation of current sheets and lead to the Ohmic dissipation of the jets' electromagnetic energy \citep{bodo2022}. This central role of resistivity in the generation of dissipative processes in relativistic plasma outflows highlights the necessity for a detailed understanding of the large scale properties of resistive relativistic jets.
	
	In this work, we introduce a new paraxial formalism for the description of relativistic spine jets. This formalism is valid both in the ideal MHD and resistive regimes, and is a modification of the formalisms used in \cite{chantry2018} and \cite{anastasiadis2024}. Similarly to the formalism of \cite{anastasiadis2024}, our formalism allows the use of a polytropic equation of state (EoS). 
	
	The rest of this paper is structured as follows. In Sec. \ref{sec:2}, we present a brief overview of the theory of relativistic, steady-state, and axisymmetric, ideal and resistive MHD jets. We also briefly discuss the properties of the variable adiabatic index EoS we use. Section \ref{sec:3} is dedicated to the description of the paraxial formalism we use in this work to obtain solutions that describe both ideal and resistive MHD relativistic jets. We present an analytical examination of the effects that nonzero electrical resistivity introduces in resistive jets, as well as a discussion on the singular points of the system of equations we derived through our paraxial formalism, which correspond to the sonic and Alfvén critical surfaces. In Sec. \ref{sec:4}, we present three ideal MHD solutions with correct crossing of the Alfvén critical surface. These three ideal MHD jets have asymptotic Lorentz factors typical of jets launched by accreting XRB systems, AGN and during GRBs. Resistive MHD solutions featuring the same boundary conditions at the Alfvén surface as the ideal MHD solutions are also presented, and are compared to their ideal MHD counterparts. This comparison highlights the ways in which Ohmic dissipation impacts the acceleration and collimation mechanisms, as well as the topology of characteristic lines on the poloidal plane and the geometry of the magnetic field. Our results are summarized and discussed in Sec. \ref{sec:5}.
	
	\section{Magnetohydrodynamics of stationary and axisymmetric relativistic jets}\label{sec:2}
	\subsection{Equations of motion and Maxwell's equations in Schwarzschild spacetime}
	We adopt the assumption of time-independent ($\partial_{t} = 0$) and axisymmetric ($\partial_{\phi} = 0$) relativistic outflows in Schwarzschild spacetime, which is characterized by the diagonal metric tensor with components
	\begin{equation}
		g_{tt} = -(1-\dfrac{r_{S}}{r})\, , g_{rr} = -\dfrac{1}{g_{tt}}\, , g_{\theta\theta} = r^{2}\, ,g_{\phi\phi} = r^{2}\sin^{2}{\theta}\, .
	\end{equation}
	$r_{S} = 2GM/c^{2}$ is the Schwarzschild radius of the central black hole, where $G$, $c$, are the gravitational constant and the speed of light, and $M$ is the black hole's mass. Hereafter, we use $h_{t}$ to denote the lapse function of the Schwarzschild metric
	\begin{equation}
		h_{t}(r) = \sqrt{-g_{tt}(r)} = \sqrt{1 - \dfrac{r_{S}}{r}}\, .
	\end{equation}
	
	In this framework, the equations of general-relativistic resistive MHD (henceforth GR-RMHD) are obtained by the following conservation laws \citep{anile1989}
	\begin{equation} \label{covariant}
		N^{\mu}_{;\mu} = 0, \quad T^{\mu\nu}_{;\nu} = 0, \quad \prescript{\star}{}{F}^{\mu\nu}_{;\mu} = 0, \quad F^{\mu\nu}_{;\mu} + 4\pi J^{\nu} = 0\, .
	\end{equation}
	The tensors appearing in the above relations are the mass density four-current, $N^{\mu}$, the four-current density, $J^{\mu}$, the energy-momentum tensor, $T^{\mu\nu}$, and the electromagnetic tensor, $F^{\mu\nu}$, and its Hodge dual, $\prescript{\star}{}{F}^{\mu\nu}$.
	We perform our analysis in the frame of an Eulerian static observer with unitary four-velocity $n^{\mu} = (1/h_{t},0,0,0)$, and express all spatial three-vectors in the spatial orthonormal basis $\{\bm{\hat{r}}, \bm{\hat{\theta}}, \bm{\hat{\phi}}\}$. The $3+1$ decomposition \citep{gourgoulhon2012} of the covariant laws in Eq. \ref{covariant} provides the familiar equations of mass conservation
	\begin{equation}\label{continuity}
		\bm{\nabla}\cdot(h_{t}\varGamma\rho\bm{v}) = 0\, , 
	\end{equation}
	momentum conservation
	\begin{equation}\label{momentum}
		\varGamma\rho\bm{v}\cdot\bm{\nabla}\left(\xi\varGamma\bm{v}\right) = -\bm{\nabla}P + \dfrac{J^{0}\bm{E} + \bm{J}\cross\bm{B}}{c} -\varGamma^{2}\xi\rho c^{2}\bm{\nabla}\ln{h_{t}}\, ,
	\end{equation}
	and entropy conservation (First Law of Thermodynamics)
	\begin{equation}\label{energy}
		\varGamma \rho c^{2}\bm{v}\cdot\bm{\nabla}\xi - \varGamma \bm{v}\cdot\bm{\nabla}P = Q\, ,
	\end{equation}
	where $\rho$, $P$, are the plasma's rest mass density and thermal pressure, and $\xi$ its specific enthalpy, which for polytropic relativistic flows is a function of the plasma's dimensionless temperature
	\begin{equation}
		\Theta = \dfrac{P}{\rho c^{2}}\, .
	\end{equation}
	$\bm{v}$ is the flow three-velocity and $\varGamma$ the associated Lorentz factor
	\begin{equation}
		\varGamma = \dfrac{1}{\sqrt{1 - \dfrac{v^{2}}{c^{2}}}}\, .
	\end{equation}
	Finally, $\bm{B}$, $\bm{E}$, $\bm{J}$, and $J^{0}/c$ are the magnetic field, electric field, current density, and charge density in the Eulerian observer frame. The source term, $Q$, which appears in the entropy conservation equation
	\begin{equation}\label{dissipationj}
		Q = \varGamma\bm{J}\cdot\left(\bm{E} + \dfrac{\bm{v}\cross\bm{B}}{c}\right) - \varGamma J^{0}\dfrac{\bm{v}\cdot\bm{E}}{c}\, ,
	\end{equation}
	is the electromagnetic energy per unit volume per unit time dissipated through Ohmic dissipation. $Q$ acts as a source of entropy for the fluid. 
	
	The electromagnetic field is governed by Maxwell's equations for stationary and axisymmetric electromagnetic fields in Schwarzschild spacetime \citep{mcdonald1982}
	\begin{equation}\label{flux}
		\bm{\nabla}\cdot\bm{B} = 0\, ,
	\end{equation}
	\begin{equation}\label{gauss}
		\bm{\nabla}\cdot\bm{E} = \dfrac{4\pi}{c} J^{0}\, ,
	\end{equation}
	\begin{equation}\label{faraday}
		\bm{\nabla}\cross\left(h_{t}\bm{E}\right) = 0\, ,
	\end{equation}
	\begin{equation}\label{ampere}
		\dfrac{\bm{\nabla}\cross\left(h_{t}\bm{B}\right)}{h_{t}} = \dfrac{4\pi}{c}\bm{J}\, .
	\end{equation}
	
	The solenoidal condition (Eq. \ref{flux}) allows us to define the magnetic flux function, $A$, which determines the magnetic field as
	\begin{equation}\label{mag}
		\bm{B} = \dfrac{\bm{\nabla}A\cross\bm{\hat{\phi}}}{\varpi} + B^{\hat{\phi}}\bm{\hat{\phi}}\, ,
	\end{equation}
	where $\varpi = r\sin{\theta}$ is the cylindrical radius from the jet's symmetry axis.
	Similarly, the continuity equation (Eq. \ref{continuity}) implies that there exists a mass flux, $\varPsi$, such that
	\begin{equation}\label{vel}
		\bm{u} = \dfrac{\bm{\nabla}\varPsi\cross\bm{\hat{\phi}}}{4\pi\rho\varpi} + u^{\hat{\phi}}\bm{\hat{\phi}}\, ,
	\end{equation}
	with $\bm{u} = h_{t}\varGamma\bm{v}$. Additionally, due to the assumption of axisymmetry, Faraday's law of induction (Eq. \ref{faraday}) allows us to define the electric potential function $\varPhi$ as \citep{thorne1982}
	\begin{equation}
		\bm{E} = -\dfrac{\bm{\nabla}(h_{t}\varPhi)}{h_{t}}\, .
	\end{equation}
	According to this last equation the electric field has no toroidal component due to the assumption of axisymmetry. To simplify the following analysis, we absorb the lapse function, $h_{t}(r)$ into the electric potential function, $\varPhi(r,\theta)$. The definition of the electric field then becomes
	\begin{equation}\label{el}
		\bm{E} = -\dfrac{\bm{\nabla}\varPhi}{h_{t}}\, .
	\end{equation}
	
	\subsection{Ideal MHD axisymmetric jets}
	Ideal MHD is built upon the assumption that plasma is a perfect conductor with infinite electrical conductivity. This assumption has considerable consequences for the properties of steady-state and axisymmetric outflows, which are summarized in the ideal MHD expression for Ohm's law, also referred to as the "frozen-in" condition
	\begin{equation}\label{idealohm}
		\bm{E} = -\dfrac{\bm{v}\cross\bm{B}}{c}\, .
	\end{equation}
	The first and most immediate consequence concerns the geometry of the electromagnetic field. According to Eq. \ref{idealohm}, the electric field is at every point in the outflow perpendicular to both the poloidal magnetic field and flow velocity. Secondly, the combination of Eq. \ref{idealohm} with the assumption of axisymmetry leads to the coincidence of the constant magnetic flux and constant mass flux lines on the poloidal plane. Due to axisymmetry, $E^{\hat{\phi}} = 0$, and by using Eqs. \ref{mag}, \ref{vel} and \ref{idealohm}, it is trivial to show that the mass flux function is a function of the magnetic flux, $\varPsi = \varPsi(A)$, and so remains constant along the magnetic flux lines. A third consequence of the assumption of infinite conductivity is the fact that the poloidal magnetic flux lines are also contours of the electric potential, and so $\varPhi = \varPhi(A)$. 
	
	There are additional quantities which remain constant along magnetic flux lines in steady and axisymmetric ideal MHD outflows, known as the ideal MHD integrals or constants of motion \citep{beskin2010}. These are
	\begin{equation}
		\Psi_{A} = \derivative{\varPsi}{A}\, ,
	\end{equation}
	\begin{equation}
		\Omega(A) = c\derivative{\varPhi}{A}= \dfrac{h_{t}v^{\hat{\phi}}}{\varpi} - \dfrac{\Psi_{A}B^{\hat{\phi}}}{4\pi\varGamma\rho\varpi}\, ,
	\end{equation}
	which is the angular velocity of the poloidal magnetic field lines,
	\begin{equation}
		\mu(A) = \xi h_{t}\varGamma - 1 - \dfrac{h_{t}\varpi\Omega(A) B^{\hat{\phi}}}{\Psi_{A}c^{2}}\, ,
	\end{equation}
	which expresses the total energy flux-to-mass flux ratio, and
	\begin{equation}
		L(A) = \xi\varGamma v^{\hat{\phi}}\varpi - \dfrac{h_{t}\varpi B^{\hat{\phi}}}{\Psi_{A}}\, ,
	\end{equation}
	the total angular momentum-to-mass flux ratio.
	
	The total energy flux-to-mass flux ratio contains two terms that correspond to the two different contributions to the total energy flux. The first term expresses the normalized energy flux of the fluid or matter
	\begin{equation}
		\mu_{M} = -\dfrac{\xi U_{t}}{c} - 1 = \xi h_{t}\varGamma - 1\, ,
	\end{equation}
	while the second term is the normalized electromagnetic energy flux of the jet
	\begin{equation}
		\mu_{EM} = - \dfrac{h_{t}\varpi\Omega(A) B^{\hat{\phi}}}{\Psi_{A}c^{2}}\, .
	\end{equation}
	$U_{t}$ is the temporal component of the outflow's covariant four velocity $U_{\mu}$
	\begin{equation}
		-U_{t} = h_{t}\varGamma c.
	\end{equation}
	Far from the black hole ($r \gg r_{S}$), the spacetime becomes asymptotically flat, and so $-U_{t} = \varGamma c$.

	The last constant of motion for steady-state and axisymmetric ideal MHD outflows is the specific entropy, $K(A)$, or entropy per particle. The conservation of the specific entropy along the poloidal magnetic flux lines is a consequence of the entropy conservation equation, Eq. \ref{energy}. In ideal MHD, the source term on the right hand side of Eq. \ref{energy} vanishes. The absence of Ohmic dissipation in ideal MHD outflows renders their specific entropy a constant of motion.
	
	With the magnetic field determined by Eq. \ref{mag}, the flow velocity is given by
	\begin{equation}
		h_{t}\varGamma\bm{v} = \dfrac{\Psi_{A}}{4\pi\rho}\bm{B} + \Omega(A)\varpi\bm{\hat{\phi}}\, ,
	\end{equation}
	and the electric field is written as
	\begin{equation}\label{electric}
		\bm{E} = -\dfrac{\Omega(A)}{ch_{t}}\bm{\nabla}A\, .
	\end{equation}
	Finally, the poloidal component of the Poynting flux vector, which is defined as
	\begin{equation}\label{poynting}
		\bm{S} = \dfrac{c}{4\pi}\bm{E}\cross\bm{B}\, ,
	\end{equation}
	is also aligned with the poloidal magnetic field lines. This can be easily shown by using Eqs. \ref{mag} and \ref{electric} for the magnetic and electric fields in Eq. \ref{poynting}, which yields the following expression for the poloidal component of the Poynting flux vector
	\begin{equation}
		\bm{S}_{p} = -\dfrac{\Omega(A)B^{\hat{\phi}}}{4\pi h_{t}}\left(\bm{\nabla}A\cross\bm{\hat{\phi}}\right)\, .
	\end{equation}
	
	\subsection{The relativistic generalization of Ohm's Law}
	The closure of the system of equations of GR-RMHD is achieved by the inclusion of the relativistic generalization of Ohm's law, which in the general case of finite electrical conductivity or nonzero resistivity is covariantly expressed as
	\begin{equation}
		J^{\mu} + J_{\nu}\dfrac{U^{\mu}U^{\nu}}{c^{2}} = \sigma\dfrac{F^{\mu\nu}U_{\nu}}{c}\, , \label{ohm}
	\end{equation}
	with $\sigma$ the plasma's electrical conductivity. Combining the $\mu = 0$ and $\mu \neq 0$ components of Eq. \ref{ohm} provides the following closure relation for the current density \citep{komissarov2007}
	\begin{equation} \label{Ohm}
		\bm{J} = \dfrac{J^{0}\bm{v}}{c} + \sigma\varGamma\left(\bm{E} + \dfrac{\bm{v}\cross\bm{B}}{c} - \dfrac{\bm{E}\cdot\bm{v}}{c^{2}}\bm{v}\right)\, . 
	\end{equation}
	This last relation, which is valid in both flat and curved spacetimes \citep{bucciantini2013, ripperda2019}, allows us to identify the two components of the current density, the convection current
	\begin{equation}\label{convection}
		\bm{J}_{conv} = \dfrac{J^{0}\bm{v}}{c}\, ,
	\end{equation}
	and the conduction current 
	\begin{equation}\label{conduction}
		\bm{J}_{cond} = \sigma\varGamma\left(\bm{E} + \dfrac{\bm{v}\cross\bm{B}}{c} - \dfrac{\bm{E}\cdot\bm{v}}{c^{2}}\bm{v}\right)\, .
	\end{equation}
	In the infinite conductivity limit ($\sigma\rightarrow\infty$), Eq. \ref{ohm} provides the ideal MHD expression for Ohm's law, Eq. \ref{idealohm}.
	
	In the outflow's comoving frame, the $\mu\neq 0$ component of Eq. \ref{ohm} provides the following relation between the comoving electric field, $\bm{E}_{co}$ and the comoving current density
	\begin{equation}\label{comovingohm}
		\bm{J}_{co} = \sigma\bm{E}_{co}\, ,
	\end{equation}
	which coincides with the expression for Ohm's law in the Newtonian limit. The comoving electric field, $\bm{E}_{co}$, is the source of Ohmic dissipation and all nonideal phenomena which emerge in resistive MHD outflows.
	
	In resistive axisymmetric MHD outflows, the plasma's conductivity is assumed to be finite and so the "frozen-in" condition, Eq. \ref{idealohm} does not hold. The electric field of the outflow is determined by the more general form of Ohm's law, Eq. \ref{Ohm}. As a result, it can have components parallel to both the poloidal magnetic flux lines and mass flux lines or streamlines, which generally do not coincide in resistive outflows. Since the existence of an electric field component along the magnetic flux lines is allowed in resistive MHD outflows, the magnetic flux lines are not contours of the electric potential either. Due to the altered geometry of the electric field and the emergence of Ohmic dissipation, there are in general no conserved quantities along either poloidal magnetic flux lines or streamlines, apart from the magnetic flux and the mass flux respectively. As such, the quantities introduced as the ideal MHD integrals, $\Psi_{A}, \Omega, \mu, L$, are not constant along magnetic flux lines in the resistive MHD regime. Thus, the self-similar modeling of steady-state and axisymmetric resistive MHD outflows differs greatly from that of their ideal MHD counterparts, as one cannot make use of constants of motion to reduce the system of equations. 
	
	\subsection{The equation of state}
	The paraxial formalism we introduce in this work allows for the modeling of polytropic MHD outflows, thus avoiding a known caveat of meridionally self-similar jet models, which require the external heating of the jet \citep{sauty1994, anastasiadis2024}. The equation of state defines the specific enthalpy as a function of the plasma's dimensionless temperature, $\Theta$. In this work, we assume the variable adiabatic index EoS proposed by \cite{ryu2006}
	\begin{equation}
		\xi(\Theta) = 2\dfrac{6\Theta^{2} + 4\Theta + 1}{3\Theta + 2}\, .
	\end{equation}
	This EoS defines the plasma's specific enthalpy as a simple analytical function of its dimensionless temperature, $\Theta$, yet satisfies Taub's fundamental inequality \citep{taub1948}
	\begin{equation}
		(\xi(\Theta) - \Theta)(\xi(\Theta) - 4\Theta) \geq 1 \, ,
	\end{equation}  
	and is a highly accurate approximation of the Synge EoS for single-species relativistic fluids \citep{synge1957}
	\begin{equation}
		\xi(\Theta) = \dfrac{K_{3}(1/\Theta)}{K_{2}(1/\Theta)}\, ,
	\end{equation}
	also independently derived by \cite{chandrasekhar1939} and \cite{cox1968}, who obtained equivalent expressions for the specific enthalpy, $\xi(\Theta)$ (see also Appendix C of \citealt{vyas2015}). The functions $K_{3}(1/\Theta)$, $K_{2}(1/\Theta)$ in the previous expression are the modified Bessel functions of the second kind, of third and second order, respectively. 
	
	The effective adiabatic index, $\gamma_{eff}$, predicted by the \cite{ryu2006} EoS is
	\begin{equation}
		\gamma_{eff} = \dfrac{h - 1}{h - 1 - \Theta} = \dfrac{5 + 12\Theta}{3 + 9\Theta}\, ,
	\end{equation}
	which accurately reproduces the limits of the Synge EoS for cold ($\Theta \ll 1$) and hot ($\Theta \gg 1$) single-species relativistic fluids
	\begin{equation}\label{polytropic}
		\gamma_{eff} = \begin{cases}
			5/3\, ,  & \Theta \ll 1 \\
			4/3\, , & \Theta \gg 1
		\end{cases}\, .
	\end{equation}
	Moreover, the fluid's specific entropy can be calculated through the second law of Thermodynamics \citep{mignone2007}
	\begin{equation}\label{secondlaw}
		\mathrm{d}K = \dfrac{\mathrm{d}\xi}{\Theta} - \mathrm{d}\ln{P}\, .
	\end{equation}
	Integration of Eq. \ref{secondlaw} provides the following expression for the fluid's specific entropy
	\begin{equation}\label{entropy}
		K = K_{0}\ln{\left(\rho^{-1/3}f(\Theta)\right)}\, ,
	\end{equation}
	with $f(\Theta)$ a function of the dimensionless temperature, $\Theta$, of the form
	\begin{equation}
		f(\Theta) = \dfrac{\sqrt{3\Theta^{2} + 2\Theta}}{\exp\left((3\Theta + 2)^{-1}\right)}\, .
	\end{equation}
	
	Different choices for the equation of state can lead to significant differences in the dynamical properties of relativistic plasma flows \citep{mizuno2013}. The use of an accurate approximation of the physically consistent Synge EoS is fundamental in consistently capturing the thermodynamic behavior of relativistic MHD jets, especially in the resistive MHD regime, where Ohmic dissipation can play a crucial role in shaping their thermodynamic properties.
	
	\section{Paraxial modeling of relativistic outflows}\label{sec:3}
	
	\subsection{Paraxial self-similar formalism}\label{subsec:paraxial}
	Meridionally self-similar models are built upon the assumption that for small values of the polar angle, as measured from the jet's symmetry axis, the magnetic flux function follows the profile
	\begin{equation}
		A(r,\theta) = f(r)\sin^{2}{\theta}\, .
	\end{equation}
	All of the physical quantities which describe the jet, including the ideal MHD integrals, are then expressed as first-order expansions with respect to the dimensionless magnetic flux, $\alpha\sim\sin^{2}{\theta}$. The classical meridionally self-similar formalism, which is valid for both Newtonian and relativistic MHD jets in the vicinity of their symmetry axis, is built upon the fact that there are conserved quantities along the poloidal magnetic flux lines. In mathematical terms, this means that the quantities which describe the jets can be written as functions of the magnetic flux function. For a more detailed overview of the classical meridionally self-similar formalism for Newtonian and relativistic nonpolytropic jets, we refer the reader to \cite{sauty1994,meliani2006}, \cite{globus2014}, and \cite{chantry2022}. A formalism for the modeling of spine jets based on the angular expansion of the governing equations, which is similar to the paraxial formalism we introduce in this work, was employed in \cite{chantry2018} for nonpolytropic outflows and modified by \cite{anastasiadis2024} to include a polytropic equation of state. 
	
	In this work, we aimed to model the dynamics of resistive relativistic MHD spine jets in order to explore the impact of Ohmic or electromagnetic (EM) dissipation on their dynamical and electromagnetic properties. Due to the absence of the ideal MHD integrals in resistive MHD jets, we chose to follow a formalism similar to the classical one, albeit one that does not assume dependence of the rest of the MHD quantities on the magnetic flux function.
	
	In order to achieve the separation of variables and obtain solutions to the equations of GR-RMHD which describe stationary and axisymmetric relativistic jets, as a first step we assumed the following paraxial profiles
	\begin{itemize}
		\item $A(r,\theta) = A_{2}(r)\sin^{2}{\theta}$\, ,
		\item $\varPsi(r,\theta) = \varPsi_{2}(r)\sin^{2}{\theta}$\, ,
		\item $\varPhi(r,\theta) = \varPhi_{2}(r)\sin^{2}{\theta}$\, ,
		\item $B^{\hat{\phi}}(r,\theta) = B^{\hat{\phi}}_{1}(r)\sin{\theta}$\, ,
		\item $u^{\hat{\phi}}(r,\theta) = u^{\hat{\phi}}_{1}(r)\sin{\theta}$\, ,
		\item $\rho(r,\theta) = \rho_{0}(r) + \rho_{2}(r)\sin^{2}{\theta}$\, ,
		\item $P(r,\theta) = P_{0}(r) + P_{2}(r)\sin^{2}{\theta}$\, ,
	\end{itemize}
	which fully describe the dynamics of a jet. The magnetic field is normalized to the magnitude of its on-axis poloidal component at the Alfvén surface, $B_{A}$. The velocity, $v$, and length are normalized to the respective characteristic scalings that appear naturally in the equations, that is the speed of light, $c$, and Schwarzschild radius, $r_{S}$. The scalings for the rest of the quantities can be derived from these three fundamental ones.
	
	In the resistive MHD regime, the paraxial resistivity profile, defined as $\eta = \dfrac{1}{4\pi\sigma}$, is assumed to be
	\begin{equation}
		\eta(r,\theta) = \eta_{2}(r)\sin^{2}{\theta}\, .
	\end{equation}
	Generally, any function $f(\theta)$ whose Taylor expansion $\tilde{f}(\theta)$ for small $\theta$ satisfies
	\begin{equation}
		\tilde{f}(\theta) \simeq \theta^{2} + \order{\theta^{3}}\, ,
	\end{equation}
	would be a suitable choice for the resistivity profile, as only this type of function guarantees the mathematical consistency of the paraxial model developed in this work. There is a clear reason for this restriction to the form of the resistivity. Let us consider Ohm's Law in the jet's comoving frame (Eq. \ref{comovingohm}), where the expansions of the radial components of $\bm{J}_{co}$ and $\bm{E}_{co}$ are
	\begin{equation}
		\tilde{J}_{co}^{\hat{r}} = J_{co,0}^{\hat{r}}(r) + J_{co,2}^{\hat{r}}(r)\theta^{2}\, ,
	\end{equation}
	\begin{equation}
		\tilde{E}_{co}^{\hat{r}} = E_{co,2}^{\hat{r}}(r)\theta^{2}\, .
	\end{equation}
	The radial component of the comoving current is generally nonzero along the jet axis, while the respective component of $\bm{E}_{co}$ is equal to zero for $\theta = 0$. The consistency of Ohm's Law along the jet axis and for small values of $\theta$ is ensured only if the resistivity is zero along the axis and proportional to $\theta^{2}$ in its vicinity. This demand for the form of the resistivity profile stems from the dependence of the electric potential on the polar angle. In this work, we deliberately set the electric potential equal to zero along the jet axis, as in ideal MHD axisymmetric outflows. This choice was made so as to obtain resistive solutions for axisymmetric spine jets whose general characteristics do not deviate significantly from those obtained through ideal MHD.
	
	The profiles presented previously are then substituted into the equations of motion, which are then expanded with respect to the polar angle $\theta$ up to second-order terms. The Taylor expansion of the equations of energy and $\hat{r}$-momentum  conservation and, in the resistive regime, of the $\hat{r}$ component of Ohm's law with respect to $\theta$ provide polynomials of the form
	\begin{equation}
		\mathcal{F}_{0}(r) + \mathcal{F}_{2}(r)\theta^{2} = 0\, .
	\end{equation}
	The polynomials obtained from the expansions of the $\hat{\theta},\, \hat{\phi}$-momentum conservation, $\hat{\phi}$-Faraday, (in the ideal MHD limit), and $\hat{\theta}$-Ohm equations take the form
	\begin{equation}
		\mathcal{F}_{1}(r)\theta = 0\, .
	\end{equation}
	The coefficients $\mathcal{F}_{i}(r)$ that appear in the above polynomials comprise a system of eight ODEs and one algebraic equation. The algebraic equation is the coefficient of the polynomial obtained from the expansion of the $\hat{\theta}$ component of Ohm's law, and determines the radial profile of the electric potential, $\varPhi_{2}(r)$, as
	\begin{equation}\label{phi2}
		\varPhi_{2}(r) = \dfrac{\Omega(r)A_{2}(r)}{c}\, .
	\end{equation}
	The polynomial obtained from the expansion of the $\hat{\phi}$ component of Ohm's law provides the following ordinary differential equation (ODE)
	\begin{equation}
		\derivative{\ln{A_{2}(r)}}{r} - \derivative{\ln{\varPsi_{2}(r)}}{r} = 0\, .
	\end{equation}
	This ODE has the simple solution
	\begin{equation} \label{Psi}
		\varPsi_{2}(r) = \Psi_{A}A_{2}(r)\, .
	\end{equation}
	According to this relation, close to their axes, the mass flux function is proportional to the magnetic flux function in resistive MHD jets, and so the magnetic flux lines and the streamlines coincide, as in ideal MHD jets. Deviations in the shape of the streamlines from that of the magnetic flux lines are found in terms of $\order{\theta^{3}}$ in the expansion of $\hat{\phi}-$ Ohm, which we do not take into account.
	
	Having utilized the algebraic equation obtained from the expansion of $\hat{\theta}-$Ohm to determine $\varPhi_{2}(r)$, and the ODE obtained from $\hat{\phi}-$Ohm to acquire Eq. \ref{Psi}, we are left with a system of seven nonlinear ODEs. These seven ODEs are second order ODEs, as they contain the second derivative of $A_{2}(r)$. They can be simplified to first order ODEs by defining the parameter
	\begin{equation}\label{expansionfactor}
		F(r) = \derivative{\ln{A_{2}(r)}}{\ln{r}}\, ,
	\end{equation}
	which is called the expansion factor \citep{globus2014}. The expansion factor is an important parameter, which contains valuable information regarding the shape of the poloidal magnetic flux lines \citep{vlahakis1999}.
	
	The value of the expansion factor, $F$, determines the shape of the field lines. For $F < 0$, the field lines turn toward the equatorial plane, while for $F = 0$, the poloidal magnetic field, $B_{p}$, assumes a purely radial (monopole) configuration. For $ 0 < F < 2$, the field lines obtain a general paraboloidal shape with $F = 2$ corresponding to a cylindrical configuration. For $F > 2$, the field lines turn toward the polar axis. Another, more intuitive way to measure a jet's degree of collimation is by calculating its (half)-opening angle along the full length of the jet. $\theta_{j}$ is defined as
	\begin{equation}
		\theta_{j}(r,\theta) = \arctan\left(\dfrac{\varpi_{j}(r,\theta)}{z_{j}(r,\theta)}\right)\, ,
	\end{equation}
	where $\varpi_{j}(r,\theta)$ is the cylidrical radius of the jet and $z_{j}(r,\theta)$ the height above the midplane along the jet boundary.
	
	The paraxial formalism we have employed in this work has considerable advantages. The first is the fact that it allows the use of a polytropic EoS for the description of the jets' thermodynamic properties. Another advantage is the simplified expressions for the quantities which describe both ideal MHD and resistive jets, which we obtain through the angular expansions of said quantities. This last advantage allows for a comprehensive understanding of the effects of Ohmic dissipation on the jets' dynamics, and the deviations from the ideal MHD behavior introduced by the assumption of a nonzero electrical resistivity.
	
	Additionally, thanks to the paraxial formalism we employ, the shapes of the characteristic lines on the poloidal plane are determined by simple parametric equations. For instance, the shape of the poloidal magnetic flux lines is determined by the simple parametric equation
	\begin{equation}
		\theta_{A}(r,\theta_{0}) = \sqrt{\dfrac{A_{2}(r_{0})}{A_{2}(r)}}\theta_{0}^{2}\, .
	\end{equation}
	$r_{0}$ is the radial distance of the point at which the magnetic flux line forms an angle $\theta_{0}$ with the polar axis. The streamlines and magnetic flux lines coincide and so we use these two terms interchangeably in the rest of the paper. The shape of the electric potential contours is accordingly given by
	\begin{equation}
		\theta_{\varPhi}(r,\theta_{0}) = \sqrt{\dfrac{\varPhi_{2}(r_{0})}{\varPhi_{2}(r)}}\theta_{0}^{2}\, .
	\end{equation}
	In ideal MHD jets, these two expressions coincide.
	
	The acceleration and collimation profiles of relativistic jets are shaped by the poloidal components of the forces which act on them. In particular, collimation is determined by the forces acting in the transfield direction, that is, perpendicularly to the streamlines. On the other hand, the characteristics of the outflow's acceleration are a product of the forces acting in the wind direction, along the streamlines. These two directions are defined by the two following unit vectors
	\begin{itemize}
		\item $\bm{\hat{t}} = \dfrac{\bm{\nabla}A\cross\bm{\hat{\phi}}}{|\bm{\nabla}A|}$, the unit vector along the streamlines
		\item $\bm{\hat{n}} = \dfrac{\bm{\nabla}A}{|\bm{\nabla}A|}$, the unit vector in the transfield direction.
	\end{itemize}
	The expressions for the angular expansions of these two unit vectors are given in Appendix \ref{sec:angular}.
	
	The poloidal forces of interest in this work are the thermal pressure gradient, the Lorentz force, and the force due to the gradient of the specific enthalpy along the streamlines, called the temperature force \citep{vlahakis2003}
	\begin{equation}
		\bm{f}_{P} = -\bm{\nabla}P,\, \bm{f}_{L} = \dfrac{J^{0}\bm{E} + \bm{J}\cross\bm{B}}{c},\, \bm{f}_{T} = - \varGamma^{2}\rho\left(\bm{v}\cdot\bm{\nabla}\xi\right)\bm{v}\, .
	\end{equation}
	These are the main forces responsible for shaping the acceleration and collimation profiles of ideal and resistive MHD relativistic jets in the vicinity of their axis of symmetry. For a more detailed view of all the forces acting on relativistic jets see \cite{anastasiadis2024}. As the solutions which we present in the following sections indicate, the jet's collimation profile is shaped by the transfield components of the thermal pressure gradient and of the Lorentz force
	\begin{equation}
		\bm{f}_{P\perp} = \bm{f}_{P}\cdot\bm{\hat{n}},\, \bm{f}_{L\perp} = \bm{f}_{L}\cdot\bm{\hat{n}}\, .
	\end{equation}
	On the other hand, the jets are accelerated thermally, due to the "temperature" force, which by definition is oriented along the streamlines. The expressions of the angular expansions of these forces are given in Appendix \ref{sec:angular}. As these expressions show, the "temperature" force is nonzero along the jet axis, while the Lorentz force along the streamlines is proportional to $\theta^{2}$, which renders it zero along the jet axis and comparatively small in the paraxial region. This is the reason why close to the jet axis, the thermal acceleration mechanism is dominant \citep{meliani2006}.
	
	We note that in our formalism the shape of the poloidal field lines is not prescribed, but determined self-consistently by the solution. Our system of equations contains all three components of the momentum equation. Therefore, by solving this system of equations, we simultaneously satisfy both the wind equation, which includes all of the acceleration mechanisms, and the transfield equation, which is the equation of force balance in the direction perpendicular to the field lines and determines the shape of the field lines as well as the jet's collimation profile \citep{anastasiadis2024}.
	
	\subsection{Effects of resistivity on jet dynamics}
	
	As explained in the previous subsection, in the vicinity of the jet axis the angular expansion of the electric potential function is written as
	\begin{equation}
		\varPhi(r,\theta) = \dfrac{\Omega(r)A_{2}(r)}{c}\theta^{2}\,.
	\end{equation}
	The reason behind the emergence of dissipative phenomena in resistive MHD jets is the dependency of the angular velocity of the poloidal magnetic field lines on the radial distance $r$. In ideal MHD jets, $\Omega$ is a constant.
	
	This is made clear by projecting Ohm's law expressed in the Eulerian observer frame (Eq. \ref{Ohm}) along the poloidal flux lines. The projection of Eq. \ref{Ohm} along $\bm{\hat{t}}$ close to the jet axis provides us with the following relation
	\begin{equation}\label{ohmalonglines}
		\left(\dfrac{\Psi_{A}^{2}|\bm{\nabla}A|^{2}A}{16\pi^{2}\varGamma h_{t}^{2}\rho^{2}c^{2}\varpi^{2}} - \varGamma A\right)\left(\bm{\hat{t}}\cdot\bm{\nabla}\Omega\right) = 4\pi\eta(\bm{J}c - J^{0}\bm{v})\cdot\bm{\hat{t}}\, .
	\end{equation}
	Expanding this equation up to second-order terms in $\theta$, we obtain the expression for the derivative of $\Omega(r)$ along poloidal flux lines
	\begin{equation}\label{dOmega}
		\Omega^{\prime}(r) = \dfrac{4\pi\eta_{2}(r)\varGamma_{0}(r)}{A_{2}(r)}\left(c J_{0}^{\hat{r}}(r) -  J^{0}_{0}(r)v^{\hat{r}}_{0}(r)\right)\, ,
	\end{equation}
	where $J^{0}_{0}(r)$ is the on-axis charge density (Eq. \ref{charge}) and $J_{0}^{\hat{r}}(r)$ is the on-axis radial current density (Eq. \ref{current}).
	
	The terms inside the parentheses in Eq. \ref{dOmega} are the on-axis term in the angular expansion of $J_{cond}^{\hat{r}}$. $J_{cond}^{\hat{r}}$ is the radial component of the conduction current, Eq. \ref{conduction}. We note that the expansion of Eq. \ref{ohmalonglines} provides the same expression for $\Omega^{\prime}(r)$ as the expansion of the $\hat{r}$-Ohm component. 
	
	The gradient of $\Omega(r)$ along the magnetic flux lines generates a gradient of the electric potential, $\bm{\nabla}_{A}\mathcal{\varPhi}$, along them, which is
	\begin{equation}\label{varphi}
		\bm{\nabla}_{A}\varPhi(r,\theta) = \dfrac{h_{t}(r)\Omega^{\prime}(r)A_{2} (r)}{c}\theta^{2}\, .
	\end{equation}
	As a result, the electric field acquires a component along the poloidal magnetic field. This additional component is part of the radial electric field (Eq. \ref{efield}.) The first term of the radial component is the radial electric field determined by the ideal MHD Ohm's law Eq. \ref{idealohm}. The second term appears as a consequence of the electric potential gradient along the magnetic flux lines. Due to the second term, the geometry of the electric field diverges from the one imposed on it by the "frozen-in" condition of ideal MHD. The angle between the electric and the poloidal magnetic field in resistive jets is
	\begin{equation}
		\chi_{EB}(r,\theta) = \dfrac{\pi}{2} + \dfrac{h_{t}(r)}{2}\derivative{\ln{\Omega(r)}}{\ln{r}}\theta\, .
	\end{equation}
	
	The power per unit volume dissipated through Ohmic dissipation, Eq. \ref{dissipationj}, is directly related to the potential difference along magnetic flux lines
	\begin{equation}\label{dissipation}
		Q(r,\theta) = \dfrac{\Omega^{\prime}(r)^{2}A_{2}(r)^{2}}{4\pi c^{2}\eta_{2}(r)}\theta^{2}\, .
	\end{equation} 
	From Eqs. \ref{dOmega} and \ref{varphi}, we deduce that $\bm{\nabla}_{A}\varPhi(\theta_{i}, r)$ depends linearly on the on-axis jet Lorentz factor, $\varGamma_{0}(r)$, since $u^{\hat{r}}_{0}(r)\simeq\varGamma_{0}(r)$, while the dissipated power per volume along magnetic flux lines, $Q$, is $Q\sim\varGamma_{0}(r)^{2}$. Consequently, we expect EM dissipation to be more significant in strongly relativistic outflows.
	
	Moreover, in the paraxial region, the Poynting flux, $\bm{S}(r,\theta)$, is the same as in ideal MHD jets. The modification due to the potential difference along the flux lines in resistive jets appears in third-order terms in the expansion of $\bm{S}(r,\theta)$.
	
	The shape of the electric potential contours on the poloidal plane is affected by the development of an electric potential gradient along the magnetic flux lines. Their shape is determined by the solution to the following differential equation 
	\begin{equation}
		\derivative{\ln{\theta_{\varPhi}}}{r} = -\dfrac{1}{2}\left(\derivative{\ln{A_{2}}}{r} + \derivative{\ln{\Omega}}{r}\right)\, ,
	\end{equation}
	which is
	\begin{equation}
		\theta_{\varPhi}(r,\theta_{0}) = \sqrt{\dfrac{\Omega(r_{i})}{\Omega(r)}}\theta_{A}(r,\theta_{0})\, .
	\end{equation}
	In ideal MHD jets, $\Omega$ is a constant and so the above relation reduces to 
	\begin{equation}
		\theta_{\phi}(r,\theta_{0}) = \theta_{A}(r,\theta_{0})\, ,
	\end{equation}
	which expresses the fact that the electric potential contours coincide with the magnetic flux lines.
	
	In general, apart from Ohmic dissipation, resistivity causes another nonideal phenomenon, the diffusion of the magnetic field. The significance of both of this phenomena can be estimated by defining appropriate dimensionless quantities. For the diffusion of the magnetic field, this number is the magnetic Reynolds number, $R_{m}$, defined as
	\begin{equation}
		R_{m} = \dfrac{\varGamma v L}{\eta c^{2}}\, ,
	\end{equation}
	where $L$ is the characteristic length over which the magnetic field varies significantly. For the jets studied in this paper, this length is the jet cylindrical radius $\varpi_{j}$, and so $L = \varpi_{j}$. The second dimensionless number, $R_{\beta}$ measures the relevant strength of EM dissipation. Just as the magnetic Reynolds number is defined by comparing the magnetic advection and diffusion terms in the induction equation, $R_{\beta}$ is defined by comparing the specific enthalpy derivative and the dissipation source term in the entropy conservation equation, Eq. \ref{energy}.
	The expression derived for $R_{\beta}$ from this comparison is
	\begin{equation}
		R_{\beta} = \dfrac{\varGamma\rho c^{2} v \xi}{Q\varpi_{j}}\, .
	\end{equation}
	This dimensionless number is the relativistic generalization of the dimensionless number which measures the significance of Ohmic dissipation in Newtonian jets, introduced in \cite{cemeljic2008}.

	\subsection{Critical surfaces}
	
	Self-similar solutions of steady-state and axisymmetric MHD outflows are characterized by the existence of critical surfaces. These critical surfaces appear at distances from the base of the outflow at which a component of the flow velocity along a certain direction becomes equal to the propagation velocity, along the same direction, of one of the characteristic MHD wave modes which can propagate in the outflow. In ideal MHD outflows, these characteristic wave modes are the Alfvén wave mode, and the slow and fast magnetosonic wavemodes. The component of the flow velocity which determines the position of the critical surfaces depends on the type of self-similarity, and consequently, symmetry of the system which one assumes. The  choice of a particular type of self-similarity imposes an additional symmetry upon the system, which enables the separation of variables. The velocity component along the direction in which the outflow presents no symmetry, or perpendicular to the two directions of symmetry, is the one that determines the position of the critical surfaces \citep{contopoulos1995, tsinganos1996}. For instance, in radially self-similar outflows, in which symmetry breaks along the $\hat{\theta}$ direction, it is the $\hat{\theta}$ component of the flow velocity which becomes equal to the wave propagation velocities at the critical surfaces \citep{vlahakis2000, vlahakiskonigl2003}. Correspondingly, in our paraxial formalism, the locations of the critical surfaces are determined by the $\hat{r}$ component of the flow velocity. We must note, however, that in our formalism only two critical surfaces appear in both the ideal MHD and resistive MHD system of ODEs. These correspond to the Alfvén and sonic wave modes. The reason behind this is the decoupling of the magnetosonic wave modes into the Alfvén and sonic wave modes along the jet axis ($\theta = 0$). The fastest of these two wavemodes corresponds to the fast magnetosonic wave mode. As such, a solution which crosses the critical surface associated with the fastest of these two modes is essentially trans-fast magnetosonic \citep{anastasiadis2024}.
	
	The singular points of our system of ODEs are determined by casting the system into the form
	\begin{equation}
		\derivative{Y_{i}}{r} = \dfrac{\mathcal{N}_{i}(r,\bm{Y})}{\mathcal{D}_{i}(r,\bm{Y})}\, , \label{frac}
	\end{equation}
	where $\bm{Y}$ is the column vector of the unknown variables 
	\begin{equation}
		\bm{Y} = \left(A_{2}(r) \, F(r) \, B^{\hat{\phi}}_{1}(r) \, u^{\hat{\phi}}_{1}(r) \, \rho_{0}(r) \, P_{0}(r)  \,\rho_{2}(r) \, P_{2}(r)\right)^{\mathrm{T}}\, .
	\end{equation}
	and calculating its determinant, which contains the term
	\begin{equation}
		(v^{\hat{r}}_{0}(r) - v_{s0}(r))(v^{\hat{r}}_{0}(r) - v_{A0}(r))\, .
	\end{equation}
	$v^{\hat{r}}_{0}(r),\, v_{s0}(r),\, v_{A0}(r)$ are the zeroth order terms in the angular expansions of the radial velocity, sound speed, and Alfvén velocity. The distances at which the on-axis velocity of the jet becomes equal to the sound speed and Alfvén velocity are the locations of the sonic and Alfvén critical surfaces. 
	
	At the positions of the critical surfaces, the determinant, and so the denominators, $\mathcal{D}_{i}$, become equal to zero. So, from a mathematical perspective, the critical surfaces are the singular points of the ODE system we obtained from the angular expansion of the equations of motion. At these points, the derivatives of the unknown variables, $\derivative{\bm{Y}}{r}$, become indeterminate. In order to avoid this and ensure the smooth crossing of a critical surface by a solution, we must ensure that the denominators, $\mathcal{N}_{i}$, are also zero at the position of that critical surface. By this demand, we obtain the regularity conditions which must be satisfied in order to achieve the correct crossing of the critical surface.
	
	For the solutions we present in the following sections, we chose to smoothly cross the Alfvén critical surface, in order to obtain trans-Alfvénic solutions which describe ideal and resistive relativistic jets. We ensure that the jet's on-axis magnetization is high enough that the Alfvén proper velocity is higher than the sound proper speed, so that the Alfvén critical surface is located at a larger distance from the black hole than the sonic critical surface. In order to smoothly cross the Alfvén critical surface, we employ the following algorithm. Firstly, we provide the values of the magnetic flux function $A_{2}(r)$, expansion factor, $F(r)$, toroidal magnetic field, $B^{\hat{\phi}}_{1}(r)$, on-axis dimensionless temperature, $\Theta_{0}(r)$, and off-axis correction to the rest mass density $\rho_{2}(r)$ at the position of the Alfvén critical surface, $r=r_{A}$, which we choose arbitrarily. We then obtain the value of the on-axis rest frame density $\rho_{0}(r)$ by demanding that $v^{\hat{r}}_{0}(r) = v_{A}(r)$ at the Alfvén surface. The off-axis correction to the thermal pressure $P_{2}(r)$ and azimuthal velocity $v^{\hat{\phi}}_{1}(r)$ at $r=r_{A}$ are determined by the regularity conditions. The on-axis thermal pressure is also determined at this step, through the values of $\Theta_{0}(r_{A})$ and $\rho_{0}(r_{A})$. We chose to determine a priori the value of the on-axis dimensionless temperature and not of the on-axis thermal pressure, as it is this quantity which determines the jet's specific enthalpy at the critical surface, regardless of the value of the on-axis rest mass density, which we cannot determine before solving for the roots of the system's determinant. In the next step, we apply L'Hôpital's rule to Eq. \ref{frac} and solve the resulting system of algebraic equations for the values of the derivatives of the unknown variables at the Alfvén surface, $\left.\derivative{Y_{i}}{r}\right\rvert_{r_{A}}$. The full trans-Alfvénic solution is comprised of a sub-Alfvénic and a super-Alfvénic branch. The sub-Alfvénic branch is obtained by integrating the system of ODEs in the upstream direction. The integration is initialized at $r = r_{A} - \delta$, with $\delta$ a small displacement, and terminates at the location of the sonic critical surface, located near the black hole event horizon, at a few Schwarzschild radii. The super-Alfvénic branch is integrated downstream from $r_{A} + \delta$ to infinity. The boundary conditions for each branch are the values of the system's variables at $r_{A} - \delta$ and $r_{A} + \delta$ respectively. These values are determined with respect to the values of the unknown variables at the Alfvén surface as
	\begin{equation}
		\bm{Y}_{r_{A} \pm \delta} = \bm{Y}_{r_{A}} \pm \left.\derivative{\bm{Y}}{r}\right\rvert_{r_{A}}\delta\, .
	\end{equation}
	
	The correct crossing of at least the critical surface which corresponds to the fastest wave mode is the most important technical aspect in the procedure of obtaining solutions which describe self-similar MHD outflows. This is because it ensures the disconnection of the outflow base from the conditions at infinity. Disturbances which emerge due to the interaction of the jet with the ambient medium at infinity can travel upstream with velocity equal to or less than the propagation velocity of the fastest wave mode. Thus, in solutions which smoothly cross the fastest wave mode critical surface, they are not able to reach the base of the jet and affect the conditions in that region. The fastest wave mode critical surface therefore, acts as an "event-horizon" for the outflow, as no information propagating in the upstream direction in the form of MHD waves can reach the regions of the outflow inside this surface \citep{sauty2002}.
	
	\subsection{Validity of the angular expansions}
	
	The solutions we present in the following section have been obtained by solving the ODEs derived by applying the paraxial formalism we introduced in the previous subsections. Since this formalism is based on the angular expansion of the equations of motion and of the quantities which describe the jet, we must determine the region of our solutions over which the errors in the expanded quantities is adequately small. The solutions must be truncated at the boundary of this region of validity of the angular expansions.
	
	The relevant expansion errors are calculated in the following way. Suppose $g(r,\theta)$ is a function of the quantities the profiles of which we introduced in subsection \ref{subsec:paraxial}, and describes a physical quantity related to the jet. Its Taylor expansion for small $\theta$ up to second order is the polynomial of $\theta$
	\begin{equation}
		\tilde{g}(r,\theta) \simeq \sum_{i = 0}^{2} g_{i}(r)\theta^{i}\, .
	\end{equation}
	The expansion of $g(r,\theta)$ can feature either purely odd or purely even powers of $\theta$. 
	The relative percent error, $\mathcal{R}_{g}$, of this expansion can be calculated as
	\begin{equation}
		\mathcal{R}_{g} = 100\dfrac{g_{j}(r)\theta^{j}}{\sum_{i = 0}^{2} g_{i}(r)\theta^{i}} \% \, , j > 2\, ,
	\end{equation}
	where $g_{j}(r)\theta^{j}$ is the next term in the expansion of $g(r,\theta)$ after the second-order term.
	Depending on the function $g(r,\theta)$, $j = 3$, or $j = 4$. The region of validity of a solution extends to where the relative percent error $\mathcal{R}_{g}$ is lower than a certain value, which we are free to choose.
	
	The quantity which must satisfy the above criterion is the one whose relative percent error displays the fastest increase with $\theta$. In the solutions we have obtained, there are three quantities which have comparable errors, the specific enthalpy, $\xi(r,\theta)$, the Lorentz factor, $\varGamma(r,\theta)$, and the matter component of the total energy flux-per-mass flux, $\mu_{M}(r,\theta)$. Depending on the solution, or the radial distance from the central black hole for one particular solution, any one of these three quantities can present the largest relative percent error. As such we enforce a strict criterion for the validity of the angular expansions, according to which the expansions of all three quantities must have a relative error $R_{\xi,\varGamma,\mu_{M}} < 10\%$. 
	\section{Solutions}\label{sec:4}
	
	\subsection{Ideal MHD solutions}
	
	In this section, we present three fiducial ideal MHD solutions which describe relativistic spine jets, with maximum terminal Lorentz factors $\varGamma_{\infty, max} \sim 5$ (XRBId solution), $\varGamma_{\infty, max} \sim 15$ (AGNId solution), and $\varGamma_{\infty, max} \sim 134$ (GRBId solution). The position of the Alfvén surface is set at $r_{A} = 10\, r_{S}$ for all three solutions. The solutions are characterized by on-axis magnetization $\sigma_{0}(r_{A}) = 10$ at the Alfvén surface, with the magnetization defined as 
	\begin{equation}
		\sigma(r,\theta) = \dfrac{B(r,\theta)^{2} - E(r,\theta)^{2}}{4\pi\rho(r,\theta)\xi(\rho,\theta)}\, .
	\end{equation} 
	We determine the strength of the toroidal magnetic field along the boundary streamline at the position of the Alfvén surface through the ratio 
	\begin{equation}
		b_{A} = \dfrac{|B_{p}(r_{A}, \theta_{0})|}{|B_{\phi}(r_{A}, \theta_{0})|}\, ,
	\end{equation}
	which we set equal to $40$ for all three solutions. The jet boundary is the streamline characterized by $\theta_{A}(r_{A}, \theta_{0}) = \theta_{0} = 0.2$ rads or $11.46^{\circ}$ at the Alfvén surface. Additionally, we have set the value of the off-axis density correction's radial profile $\rho_{2}(r_{A}) = 5$ and the value of the expansion factor $F(r_{A}) = -0.35$ at the Alfvén surface. These choices were made so as to minimize the relative percent errors in the angular expansions of $\xi, -U^{t}$, and $h_{t}\varGamma\xi$ for each solution. The boundary conditions and parameters of each solution are summarized in Table \ref{tab:table}.
	
	\begin{table*}
		\caption{Parameters and boundary conditions at the Alfvén surface for the three fiducial ideal MHD jet solutions.}
		\centering
		\begin{tabular}{>{\centering\arraybackslash}m{1.6cm} >{\centering\arraybackslash}m{1.6cm} >{\centering\arraybackslash}m{1.6cm} >{\centering\arraybackslash}m{1.6cm} >{\centering\arraybackslash}m{1.6cm} >{\centering\arraybackslash}m{1.6cm} >{\centering\arraybackslash}m{1.6cm} >{\centering\arraybackslash}m{1.6cm} >
				{\centering\arraybackslash}m{1.6cm}}
			\toprule
			Solution & $r_{A}\, [r_{S}]$ & $\Psi_{A}\, [\frac{B_{0}}{4\pi c}]$ & $\Theta_{0}(r_{A})$ & $\sigma_{0}(r_{A})$ & $b_{A}$ & $F(r_{A})$ & $\rho_{2}(r_{A})\, [\frac{B_{0}^{2}}{4\pi c^{2}}]$ & $\Theta_{2}(r_{A})$ \\
			\midrule
			XRBId & 10 & 5 & 0.1 & 10 & 40 & -0.35 & 5 & 3.38\\
			AGNId & 10 & 1.7 & 1 & 10 & 40 & -0.35 & 5 & 7.62\\
			GRBId & 10 & 0.55 & 10 & 10 & 40 & -0.35 & 5 & 62.84\\
			\bottomrule
		\end{tabular}
		\label{tab:table}
	\end{table*}
	
\begin{figure}
		\includegraphics[width = 0.5\textwidth]{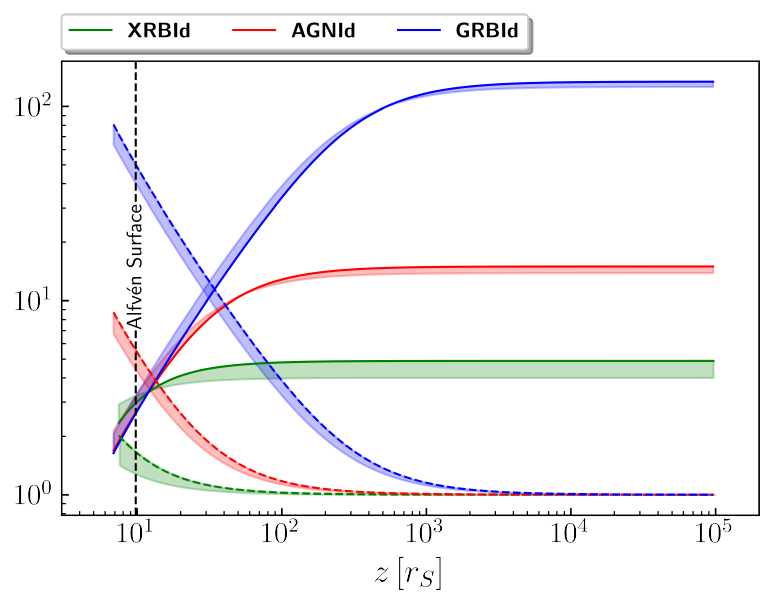}
		\caption{$-U_{t}$, (solid lines) and specific enthalpy, $\xi$, (dashed lines) for the XRBId, AGNId, and GRBId solutions. The solid and dashed lines correspond to the values of $-U_{t}$ and $\xi$ along each jet's boundary streamline. The shaded regions show the variation of each quantity as we move from the boundary streamline to the jet axis.}
		\label{avlf}
	\end{figure}

	
	The three ideal MHD solutions are accelerated thermally, with their terminal Lorentz factors determined only by their specific enthalpy, or equivalently their temperature, at their base. Jets which are hotter at their base are able to achieve higher terminal Lorentz factors, while also being accelerated over a larger distance from their base. As seen in Fig. \ref{avlf}, the GRBId solution is accelerated up to a height of $1000\, r_{S}$ over the midplane. The length of its thermal acceleration region is an order of magnitude larger than that of the AGNId solution and 2 orders of magnitude larger than in the XRBId solution, a direct consequence of its significantly higher temperature at its base. The prevalence of the thermal acceleration mechanism is an outcome of the paraxial nature of our formalism, as in the vicinity of the jets' symmetry axis the thermal acceleration mechanism is dominant \citep{meliani2006}. This feature of thermally accelerated fast jet spines is also predicted by numerical simulations of thermally dominated jets \citep{ricci2024}.
	
	Also, we observe that the Lorentz factors of the two strongly relativistic solutions (AGNId, and especially GRBId), follow the profile $\varGamma\sim z$ for a large part of their acceleration phase. Therefore, the formalism presented in this work accurately reproduces the acceleration profile of strongly relativistic thermally accelerated outflows, as predicted by the relativistic fireball model \citep{shemi1990}. The Lorentz factor of the weakly relativistic XRBId solution on the other hand displays a weaker dependence on $z$. The length over which this jet is accelerated is much shorter, with the jet reaching its terminal Lorentz factor shortly after the distance at which the Alfvén critical surface is located.
	
	\begin{figure}[H]
		\includegraphics[width = 0.5\textwidth]{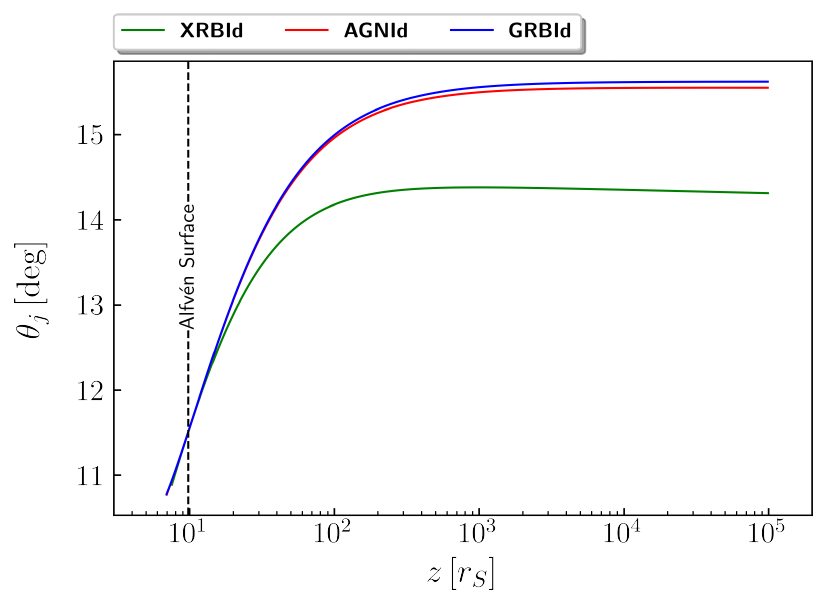}
		\caption{Jet opening angle, $\theta_{j}$, for the three fiducial ideal MHD solutions. The weakly relativistic XRBId solution displays weak collimation asymptotically, as evidenced by the decrease of its opening angle with $z$, while the AGNId and GRBId solutions feature near-radial or freely expanding streamlines. At low $z$, the opening angles of all three solutions increase, as the jets are decollimated by the positive transfield component of the Lorentz force.}
		\label{thetaj_id}
	\end{figure}
	
	Another difference observed between the two strongly relativistic solutions and the weakly relativistic one concerns their collimation. Asymptotically, at distances well beyond the Alfvén surface, the XRBId jet solution displays stronger collimation than both the AGNId and GRBId solutions. This can be observed in Fig. \ref{thetaj_id}, which shows the opening angle of the XRBId solution decreasing with $z$ at large distances from the jet base. Therefore, the weakly relativistic solution experiences stronger collimation asymptotically due to the transfield component of the Lorentz force, $f_{L\perp}$. The reason behind the weaker collimation exhibited by the AGNId and GRBId solutions is the electric decollimating force. Beyond the Alfvén surface, the AGNId and GRBId solutions reach high Lorentz factors, on the order of $10$ and $100$ respectively. As a result, the transfield electric force becomes practically equal in magnitude to the toroidal magnetic field, with the decollimating electric force almost completely cancelling out the magnetic collimating force. The streamlines of the two strongly relativistic solutions, AGNId and GRBId, obtain a radial asymptotic shape, as evidenced by the fact that their opening angles are constant at large distances from their bases.
	
\begin{figure*}
		\includegraphics[width = 1.0\textwidth]{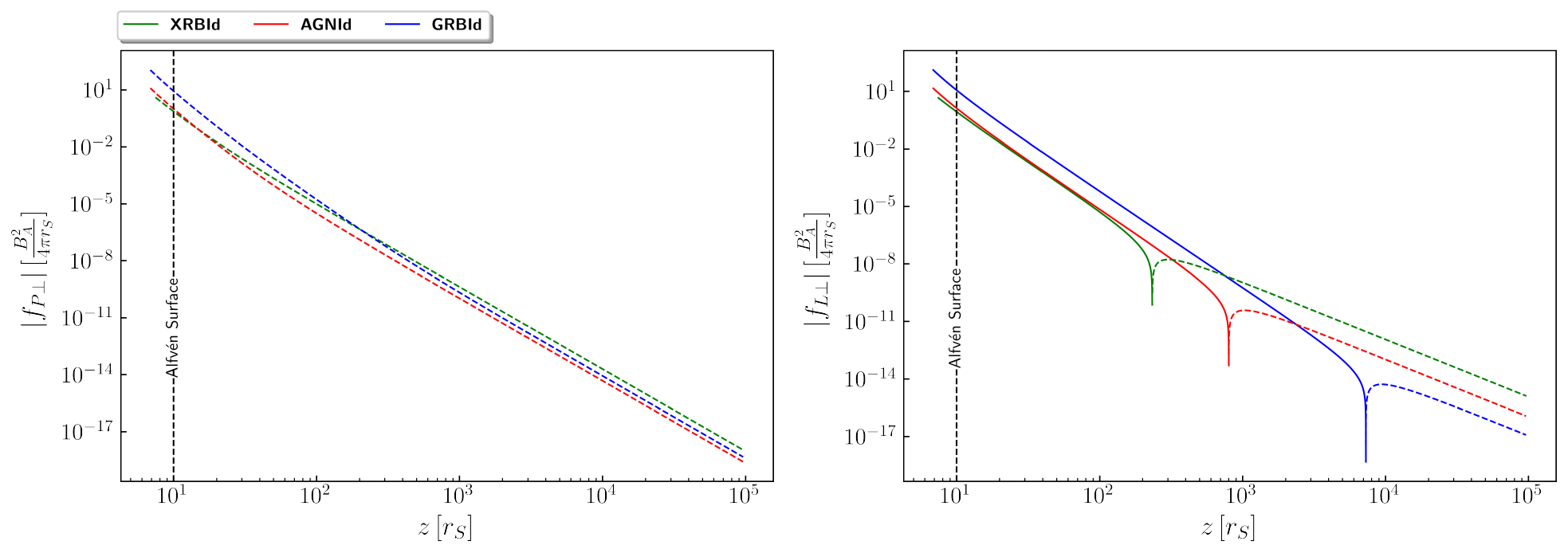}
		\caption{Magnitudes of the forces in the transfield direction for the three ideal MHD solutions along the boundary streamline. Left panel: Thermal pressure gradient. Right panel: Lorentz force. Dashed lines denote where a force is negative and thus collimates the jet.}
		\label{fn}
	\end{figure*}

	Close to the central black hole, the transfield component of the thermal pressure gradient, $f_{P\perp}$, is negative, and thus collimating, while the transfield Lorentz force, $f_{L\perp}$, which is stronger than the thermal pressure gradient, is decollimating, as shown in Fig \ref{fn}. The net effect of these two transfield forces is the decollimation of the jets. The near-radial streamline configuration of the two strongly relativistic solutions suggests that self-collimation is not sufficient to collimate relativistic jets into the conical, paraboloidal or cylindrical shapes observed in a number of cases. As numerical works have shown \citep{komissarovetal2007, rohoza2024}, the existence of an ambient environment is critical in collimating relativistic jets into the occasionally observed paraboloidal and cylindrical shapes.
	
	For our investigation of the impact of EM dissipation on the behavior of relativistic astrophysical jets we focus on the ultra-relativistic GRBId solution for reasons we explain in the next section. Nevertheless, in this section we chose to present solutions across a wide range of terminal Lorentz factors, in order to showcase the universality of our paraxial formalism.  
	
	\subsection{Resistive MHD solutions} 
	For the resistive outflow solutions presented hereafter, we chose the following resistivity profile
	\begin{equation}\label{res}
		\eta(r, \theta) = \begin{cases}
			0\,, & r < r_{\eta} \\
			\eta_{c}\tanh\left( \dfrac{r - r_{\eta}}{\sigma_{\eta}} \right)\theta^{2}\, , & r \geq r_{\eta}
		\end{cases}
	\end{equation}
	According to this profile, the resistivity begins increasing from zero at a distance $r_{\eta}$  from the base of the outflow, reaching a constant value, $\eta_{c}$. The slope of the profile is controlled by the parameter $\sigma_{\eta}$. 
	
	Dissipative effects become more prominent in strongly relativistic solutions. So, for a better illustration of the effects of EM dissipation on the properties of relativistic jets, we chose to focus on the ultra-relativistic GRBId solution in the rest of this section. We note, however, that the effects of nonzero resistivity are qualitatively the same for all solutions, regardless of their kinematic properties. As such, the conclusions drawn from the ultra-relativistic resistive solutions apply to all types of relativistic astrophysical jets. 
	
	Also, we note that only the super-Alfvénic branch of our solutions is resistive, in order to study the effects of EM dissipation away from the central object, where it can emerge due to instabilities caused by the interaction of the jet with its surrounding medium. Close to the central black hole, resistivity is more probably a result of the turbulent hot corona and accretion disk.
	
	The cases of resistive jets we considered were jets with the same boundary conditions at the Alfvén surface as the GRBId solution of the previous section, and are characterized by a nonzero electrical resistivity of the form of Eq. \ref{res}. The parameters of the resistivity profile for each resistive solution are presented in Table \ref{tab:respar}. We have obtained five resistive jet solutions in total, although in the following we concentrate our focus on solutions GRBRes1 and GRBRes2. However, we comment on the rest of the solutions when necessary, in order to highlight particular aspects of the behavior of resistive jets which are the outcome of differences in the resistivity profile used.
	
	\begin{table}
		\caption{Resistivity profile parameters for the five resistive jet solutions.}
		\centering
		\begin{tabular}{>{\centering\arraybackslash}m{1.5cm} >{\centering\arraybackslash}m{1.25cm} >{\centering\arraybackslash}m{1.25cm} >{\centering\arraybackslash}m{1.5cm}}
			\toprule
			Solution & $r_{\eta}\, [r_{S}]$ & $\sigma_{\eta}\, [r_{S}]$ & $\eta_{c}\, [r_{S}/c]$ \\
			\midrule
			GRBRes1 & 100 & 50 & 1800\\
			GRBRes2 & 100 & 50 & 1940\\
			GRBRes3 & 100 & 40 & 1780\\
			GRBRes4 & 12 & 50 & 59\\
			GRBRes5 & 2000 & 50 & $2.4\times 10^{5}$\\
			\bottomrule
		\end{tabular}
		\label{tab:respar}
	\end{table}
	
\begin{figure}
		\centering
		
		\includegraphics[width = 0.5\textwidth]{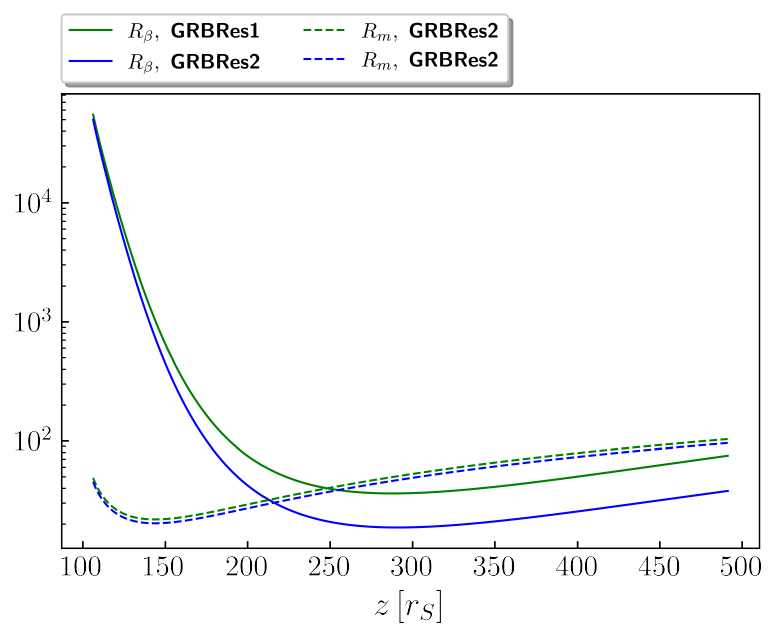}
		\caption{$R_{\beta}$ and magnetic Reynolds number, $R_{m}$, for the two resistive solutions with GRBRes1 and GRBRes2.}  
		\label{Rm}
	\end{figure}

	For both resistive solutions, GRBRes1 and GRBRes2, we find that Ohmic dissipation is slightly more significant than the diffusion of the magnetic field, as indicated by Fig. \ref{Rm}. This result is in agreement with the results of \cite{cemeljic2008} for Newtonian jets. The respective dimensionless numbers, $R_{m}$ and $R_{\beta}$, which measure the intensity of these two effects, have values much higher than unity over the entire resistive part of each solution. Thus we conclude that, in the GRBRes1 and the GRBRes2  resistive solutions, diffusion of the magnetic field is not capable of significantly affecting the geometry of the poloidal magnetic flux lines, while the adiabatic cooling of the jet is much more significant than the Ohmic dissipation of the EM field's energy. Nevertheless, Ohmic dissipation is capable of affecting the dynamics of the relativistic jets of both solutions.
	
\begin{figure}
		\centering
		
		\includegraphics[width = 0.5\textwidth]{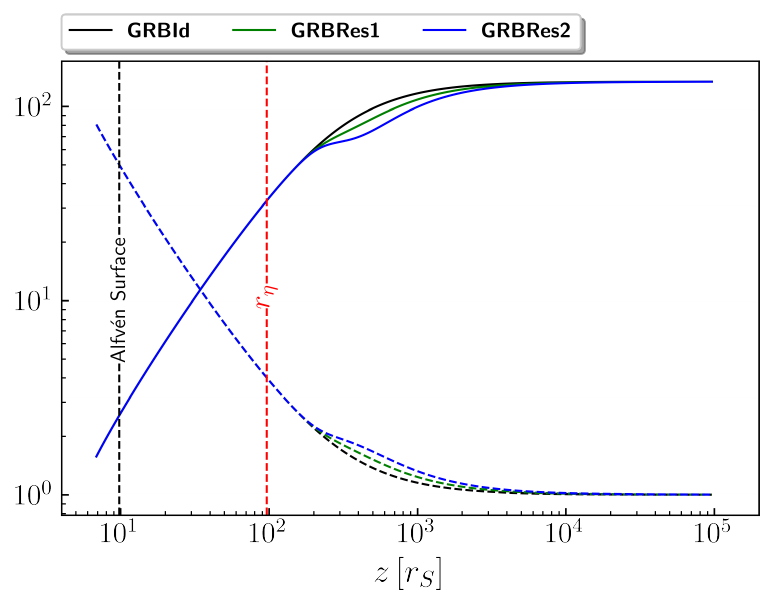}
		\caption{-$U_{t}$ (solid lines), and specific enthalpy, $\xi$, (dashed lines), for the GRBId and GRBRes1, GRBRes2 solutions. Ohmic heating softens the specific enthalpy gradient, reducing acceleration along the boundary streamline.}  
		\label{avlfr}
	\end{figure}

	As we showed in the previous section, the jets described by our solutions are thermally accelerated through the conversion of their thermal content to kinetic energy. At the same time, the most direct effect of Ohmic dissipation is the conversion EM energy into thermal energy. As such, dissipation can decrease the rate at which the off-axis part of a relativistic jet is accelerated thermally, or even decelerate it, depending on its relative significance to the adiabatic cooling along the streamlines. In both resistive solutions GRBRes1 and GRBRes2, the thermal acceleration mechanism is affected by dissipation, with the intensity of the effect increasing with the angular distance from the jets' axes. As shown in Fig. \ref{avlfr}, $-U_{t}$ along the boundary streamline of the GRBRes1 solution displays a softer dependence on $z$ compared to its ideal MHD counterpart GRBId, with the effect appearing stronger in the GRBRes2 solution, characterized by a higher value of $\eta_{c}$. The specific enthalpies of the two solutions decrease at a slower rate, too. After a certain distance from $r_{\eta}$, where dissipation becomes negligible, the two solutions are again accelerated thermally, reaching the same terminal Lorentz factor as the ideal MHD solution.
	
\begin{figure}
		\centering
		
		\includegraphics[width = 0.5\textwidth]{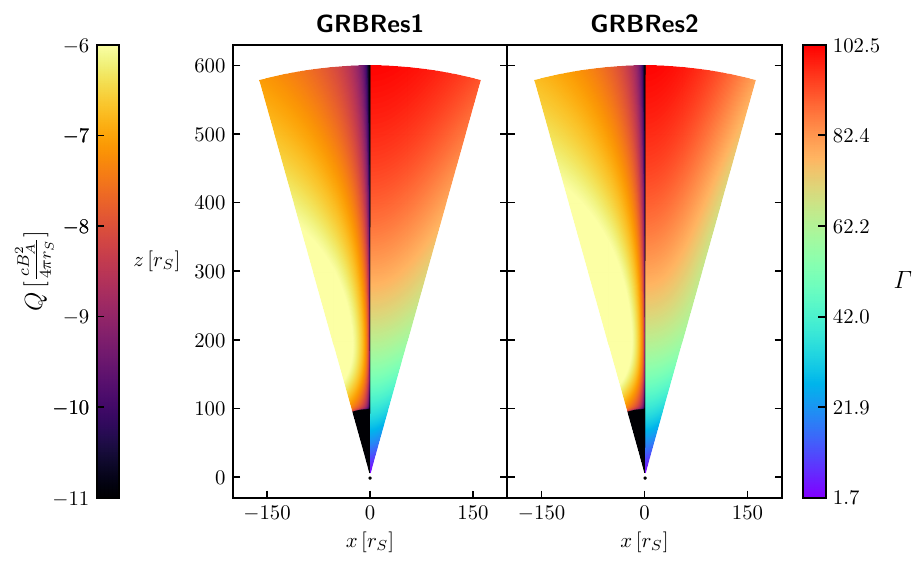}
		\caption{Dissipated power per unit volume, $Q$, in logarithmic scale and Lorentz factor, $\varGamma$, for the two resistive solutions GRBRes1, GRBRes2. The Lorentz factor colormap is normalized to the maximum Lorentz factor value of the ideal MHD solution at $r = 600\, r_{S}$, which is along the jet boundary streamline.}  
		\label{Q}
	\end{figure}

	According to Eq. \ref{dissipation}, Ohmic dissipation increases with the angular distance from the jet axis, while on the axis it is zero. This is the reason why the acceleration of the jet is more strongly affected along streamlines near the jet boundary, as shown in Fig. \ref{Q}. Along the off-axis streamlines of the resistive jets, dissipation also causes the increase of the plasma's specific entropy, $K$ (Eq. \ref{entropy}).
	
\begin{figure}
		\centering
		
		\includegraphics[width = 0.5\textwidth]{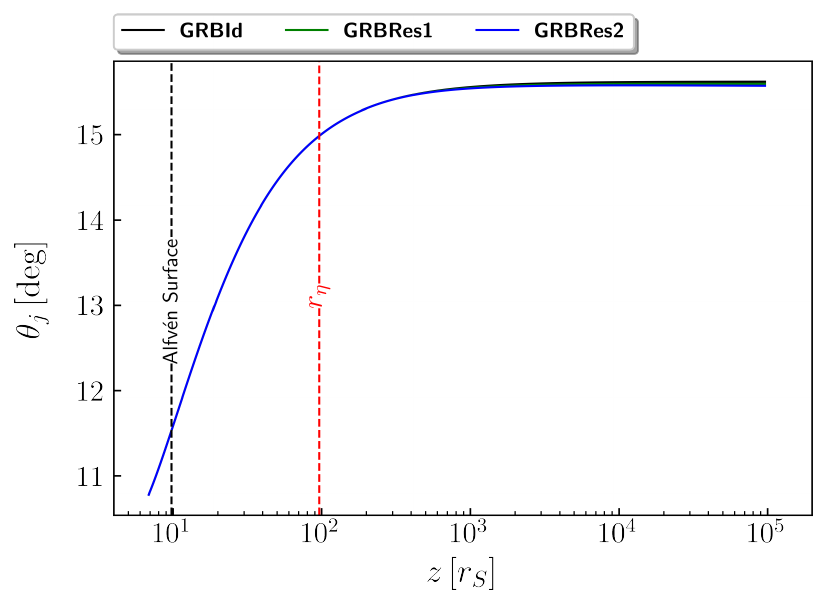}
		\caption{Jet opening angle for solutions GRBId, GRBRes1, and GRBRes2. The resistive solutions present a slightly smaller asymptotic opening angle than the ideal MHD solution, an effect of resistive MHD Ohm's law on the collimation.}  
		\label{F}
	\end{figure}

\begin{figure*}
		\centering
		
		\includegraphics[width = 1\textwidth]{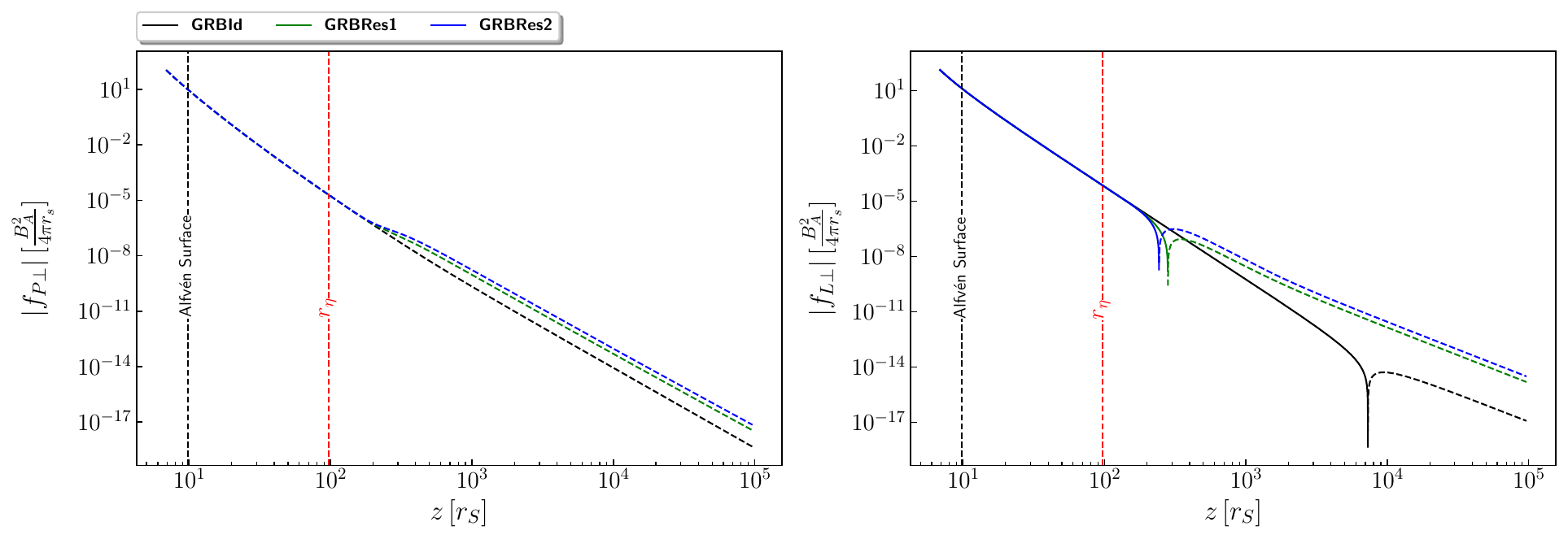}
		\caption{Forces in the transfield direction along the boundary streamlines for solutions GRBId, GRBRes1, and GRBRes2. Left panel: Thermal pressure gradient. Right panel: Lorentz force. The thermal collimating force is intensified in both resistive solutions, while the force due to the electromagnetic field becomes negative and collimates the jet shortly after $r_{\eta}$, denoted by the red vertical dashed line. The dashed parts of the curves correspond to negative values of the forces.}  
		\label{fnetac}
	\end{figure*}

	There is, however, a second, indirect way in which the plasma's nonzero resistivity can affect the kinematical properties of a relativistic jet. The resistive solutions are more collimated than the ideal MHD solution, as shown by their opening angles in Fig. \ref{F}. The stronger collimation of the resistive solutions is attributed to two factors. The first is the heating of the plasma along off-axis streamlines due to dissipation. Dissipation strengthens the collimating transfield component of the thermal pressure gradient, with this increase becoming more intense the further away we move from the jet axis. This intensifies the jet's thermal collimation. Additional, the radial current density is amplified in the resistive solutions, thereby strengthening their azimuthal magnetic fields over the dissipation region. As such, the magnetic collimation mechanism of the resistive jets is also enhanced. In Fig. \ref{fnetac} we present the magnitudes of the transfield thermal pressure gradient and Lorentz forces for the ideal MHD and the resistive solutions.
	
	In order to understand how the jets' collimation affects their acceleration, we need to keep in mind that the efficiency of their acceleration, thermal or magnetic, is connected to the shape of the streamlines on the poloidal plane. For instance, in the case of magnetic acceleration, which works away from the jet axis, the streamlines must follow a paraboloidal profile \citep{komissarovetal2007, ricci2024}. On the other hand, thermal acceleration is stronger in freely expanding jets with radial streamlines, compared to jets presenting stronger collimation. Freely expanding jets, like the ideal MHD solution presented in this section, allow for steeper gradients of the specific enthalpy along the streamlines, compared to jets which feature paraboloidal streamlines. The steeper enthalpy gradients result in a faster conversion of the plasma's thermal energy to kinetic energy along the streamlines. Additionally, collimation determines the shape of all the streamlines in the jet, indirectly affecting both the on-axis and off-axis acceleration, which is why this effect is observed when considering the Lorentz factor profile along the jet axis, and can be discerned from the off-axis Ohmic heating effect on the thermal acceleration mechanism.
	
	An easy way to compare the efficiency of the thermal acceleration mechanism between solutions is through the distance between two neighboring streamlines. In the paraxial regime, this distance is
	
	\begin{equation}
		\delta l_{\perp} \simeq r\sqrt{\dfrac{A_{2}(r_{A})}{A_{2}(r)}}\delta\theta_{i}\, .
	\end{equation}
	
	Jets for which $\delta l_{\perp}$ increases faster with the distance from their base are accelerated at a faster rate, both along the axis and along off-axis streamlines. If different trans-Alfvénic solutions are characterized by the same value of $A_{2}(r)$ on the Alfvén critical surface, like the ideal MHD and resistive MHD solutions of this section, then we can easily determine which jet is accelerated more efficiently along its axis by comparing the radial profiles of the magnetic flux function of each solution, $A_{2}(r)$. Between two solutions $a$ and $b$ with $A_{2,b}(r) > A_{2,a}(r)$ over some region past the Alfvén point, solution $a$ will be the one that displays more efficient thermal acceleration. 
	
\begin{figure}
		\centering
		
		\includegraphics[width = 0.5\textwidth]{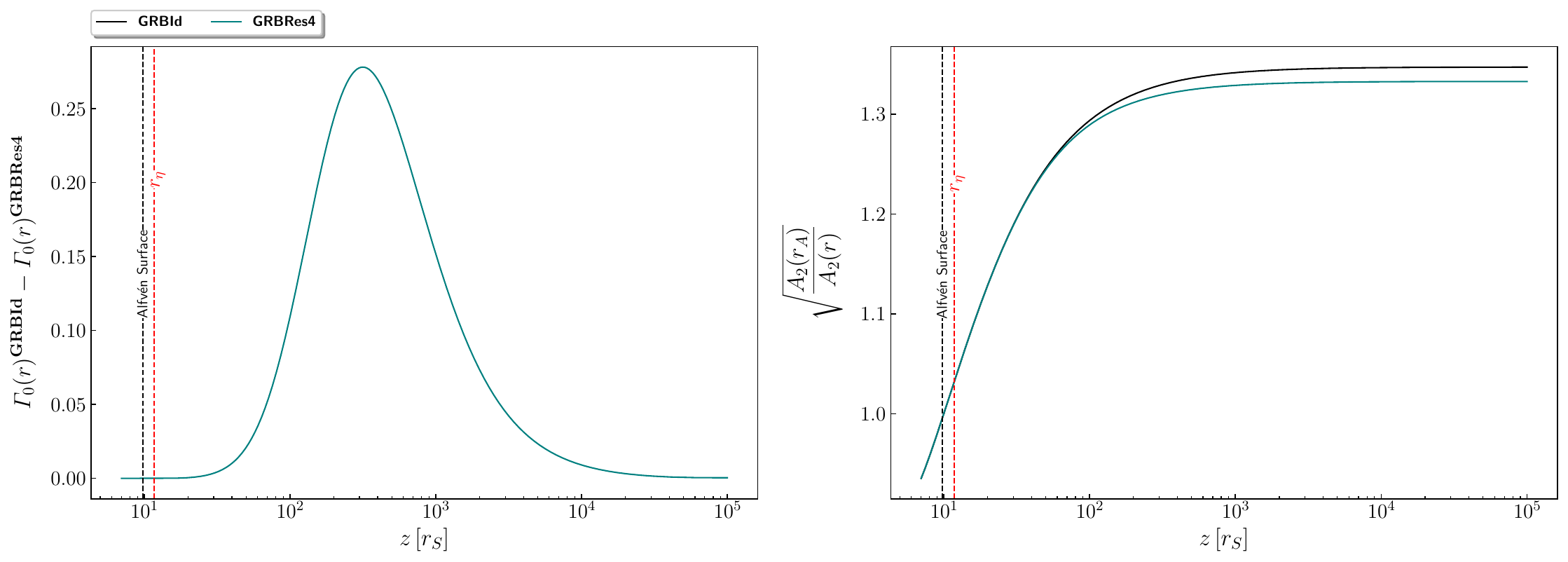}
		\caption{Effect of the stronger collimation exhibited by the resistive solution GRBRes4 on its on-axis acceleration profile. Left panel: Difference between the on-axis Lorentz factor of the ideal MHD solution, GRBId, and solution GRBRes4, with $r_{\eta} = 12\, r_{S}$. Right panel: radial part of the function which measures the distance between neighboring flux lines, $\delta l_{\perp}$, for the same two solutions.}  
		\label{onaxis}
	\end{figure}

	The modification of the acceleration profile due to the stronger collimation of the resistive solutions can be observed more easily in solutions in which we switch on the resistivity at lower heights above the jet base. In order to illustrate this, we obtained a resistive solution with a resistive profile switched on at a much shorter distance from the jet base, $r_{\eta} = 12\, r_{S}$ (solution GRBRes4). The difference between the on-axis Lorentz factors, $\varGamma_{0}(r)$ of solutions GRBId and GRBRes4 is shown in Fig. \ref{onaxis}. 
	
\begin{figure*}
		\centering
		
		\includegraphics[width = 1\textwidth]{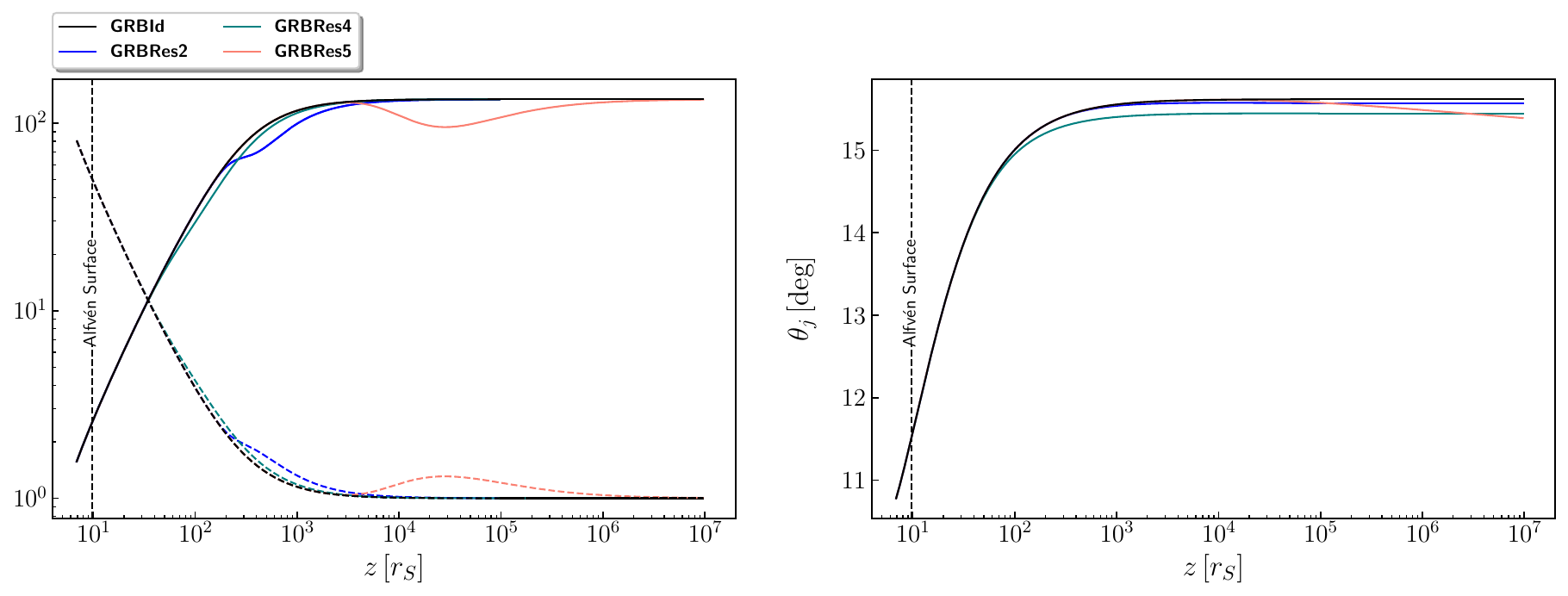}
		\caption{Effect of the location of the onset of the resistivity, $r_{\eta}$, on the acceleration and collimation of the resistive jets. Left panel: $-U_{t}$ (solid lines) and specific enthalpy, $\xi$, (dashed lines) along the jets' boundary streamlines. Right panel: jet opening angle.}  
		\label{lfs}
	\end{figure*}

	The value of $r_{\eta}$, which determines the location along the jet where the resistivity is switched on, can control the strength of Ohmic dissipation relative to the adiabatic cooling of the jet. The further away this location is from the jet base, the relative significance of Ohmic dissipation increases. The effect of this parameter on the jet acceleration and collimation is summarized in Fig. \ref{lfs}. 
	
\begin{figure*}
		\centering
		
		\includegraphics[width = 1.0\textwidth]{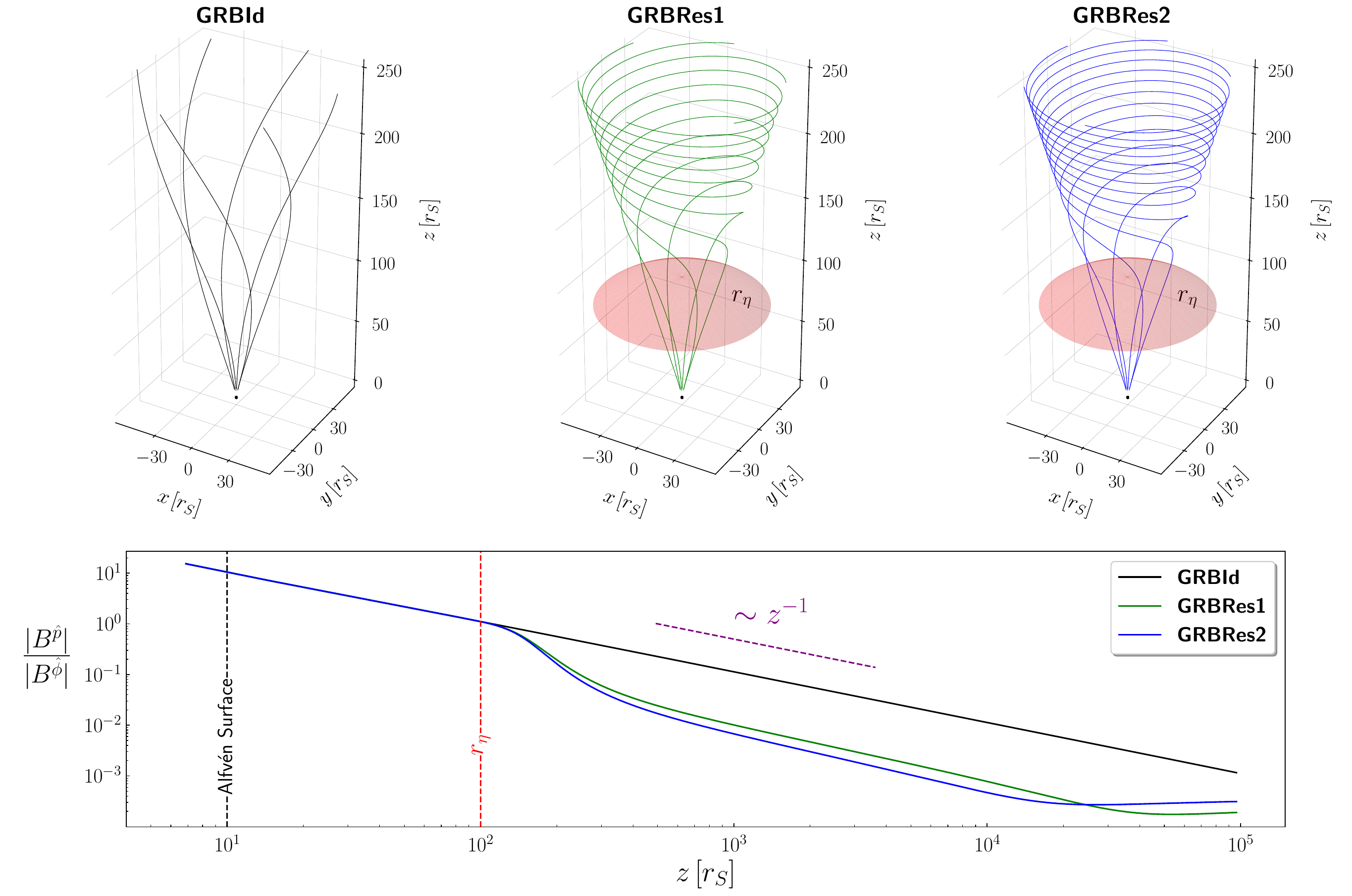}
		\caption{Structure of the 3D jet magnetic field for the ideal MHD solution, GRBId, and resistive MHD solutions, GRBRes1 and GRBRes2. In all cases the magnetic field has the form of a helix. The helix of the magnetic field in the resistive solutions is more tightly wound, with the pitch decreasing as we increase the value of $\eta_{c}$.}  
		\label{3dfield}
	\end{figure*}

	As explained previously, in resistive solutions the radial current density is amplified over the dissipation region, which in turn leads to a strengthening of the toroidal magnetic field for $r > r_{\eta}$. The differences in the magnetic field geometry between the ideal MHD solution 
	GRBId and the resistive solutions GRBRes1 and GRBRes2, which are essentially an outcome of the use of the resistive Ohm's law in the resistive jet solutions, are quite significant, as the 3D magnetic field structure shows, which is presented in Fig. \ref{3dfield}, shows. 
	
	The poloidal and toroidal magnetic field components of the ideal MHD solution, follow the scalings $|B_{p}| \sim z^{-2},\, B^{\hat{\phi}}| \sim z^{-1}$, where $z \simeq r$ close to the jet axis, as is typical for jets featuring radial (monopole) poloidal magnetic field configurations \citep{chen2021}, with the ratio of the poloidal magnetic field strength over the toroidal magnetic field strength decreasing as $z^{-1}$. In the resistive solutions, the toroidal magnetic field is amplified after the resistivity is switched on at $r = r_{\eta}$, with the ratio of the poloidal over the toroidal magnetic field following a much steeper profile over the region where Ohmic dissipation is important. At larger distances from the jet base ($r > 10^{3}\, r_{S}$), the ratio of the two magnetic field components follows a smoother profile than in the ideal MHD solution. This is due to the less steep decrease in the poloidal magnetic field strength with $z$ in the resistive solutions, which is an outcome of their stronger collimation. The increase in the strength of the toroidal magnetic field in the resistive solutions is observed over the dissipation region, up to about $10^{3}\, r_{S}$, over which the angular velocity of the magnetic field lines, $\Omega$, increases. This increase in $\Omega$ is the reason behind the toroidal magnetic field's amplification in the resistive solutions. Due to the increase in $\Omega$ over the dissipation region, the charge density, $J^{0}/c$, which is proportional to $\Omega$ (Eq. \ref{charge}), displays an increase too. As a consequence, the radial convection current (Eq. \ref{convection}), which is the dominant radial current density, is amplified. This amplification of the dominant radial current density leads to the strengthening of the toroidal magnetic field over the dissipation region. At larger distances, beyond the dissipation region, the toroidal magnetic field follows the same $|B^{\hat{\phi}}| \sim z^{-1}$ profile as in the ideal MHD solution. 
	
\begin{figure}
		\centering
		
		\includegraphics[width = 0.5\textwidth]{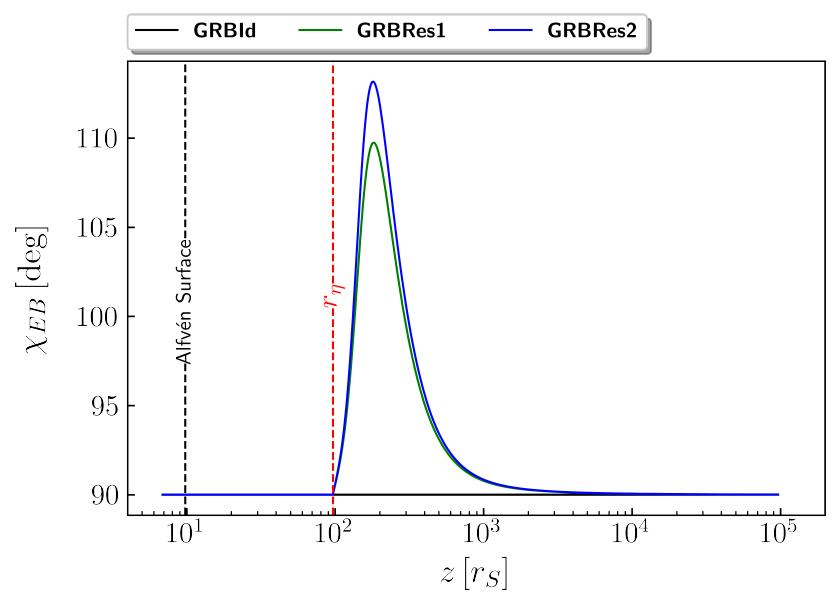}
		\caption{Angle between the electric field and the poloidal magnetic field, $\chi_{EB}$, along the boundary streamline. At $r_{\eta} = 100\, r_{S}$ where the resistivity is switched on, the electric potential gradient starts to increase along the off-axis streamlines. This generates an electric field component along the streamlines and consequently the configuration of the electric field is affected. Past $10^{3}\, r_{S}$, the gradient of $\Omega$ along the magnetic flux lines has become negligible. The electric potential gradient along the flux lines dies out and so the electric field becomes again perpendicular to the poloidal magnetic field.}  
		\label{chi}
	\end{figure}

	The topology of the electric field is affected as well, due to the generation of an electric field component parallel to the poloidal magnetic field lines in the resistive solutions. The angle between the electric field and the poloidal magnetic field, $\chi_{EB}$, displays an increase, which starts at $z = r_{\eta}$. The two fields become once again parallel after a certain distance where dissipative effects have become negligible, as portrayed in Fig. \ref{chi}. The regions of the resistive solutions where $\chi_{EB}$ deviates significantly from $\pi/2$ coincide with the dissipation regions, where effects due to resistivity and Ohmic dissipation are important.
	
	The strong modification of the helical structure of the magnetic field in regions of EM dissipation could in theory allow for the identification of such regions in astrophysical jets, by studying the polarization of their synchrotron emission. The existence of a toroidal magnetic field component and the consequent helical magnetic field structure causes a gradient of the synchrotron emission's Faraday rotation measure (RM), transversely to the jet axis, in multiple regions along the jet (\citep{hovatta2012, algaba2013}). These transverse RM gradients are observed in the emission of several AGN jets \citep{gabuzda2018}. Since resistivity strongly alters the helical configuration of the magnetic field, by amplifying its toroidal component in the jet regions where EM dissipation is significant, it is possible that EM dissipation could create observable signatures in the transverse polarization of the synchrotron emission of AGN jets.
	
\begin{figure}
		\centering
		
		\includegraphics[width = 0.5\textwidth]{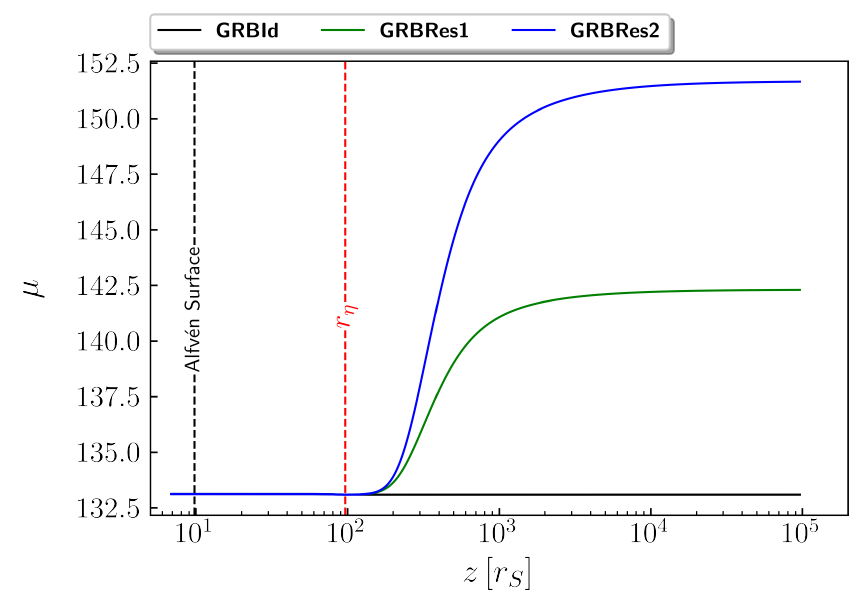}
		\caption{Total-energy-to-mass flux ratio, $\mu$, of the ideal MHD and resistive jets. The increase of $\mu$ along the boundary streamline is significant in both solutions. Solution GRBRes2, which displays stronger dissipation, also shows a larger increase in $\mu$. In both solutions, this increase is located in the region where the electric potential gradient along the flux lines is significant. At larger distances where it is negligible, $\mu$ is again constant.}  
		\label{mufig}
	\end{figure}

	The differences in the magnetic field topology between the ideal MHD and the resistive solutions, together with the electric field component parallel to the poloidal magnetic field in the resistive jets, result in a variation of the Poynting flux along the poloidal magnetic field lines. In the resistive solutions, the Poynting flux is increased along the flux lines over the dissipation region. This leads to an increase of the total-energy-to-mass flux ratio $\mu$ along the streamlines. This increase of $\mu$ along the streamlines is attributed only to its electromagnetic component, as its matter component, $\mu_{M} = -U_{t}\xi  - 1$, remains constant in both the ideal MHD and resistive jets. This is expected, as these solutions are accelerated thermally, and so any change in the specific enthalpy along the streamlines brings about the exact inverse change in $-U_{t}$ along any streamline, thus keeping $\xi h_{t}\varGamma$ constant. The variation of $\mu$ along the boundary streamlines of the two resistive jets can be seen in Fig. \ref{mufig}.
	
\begin{figure*}
		\centering
		
		\includegraphics[width = 1.0\textwidth]{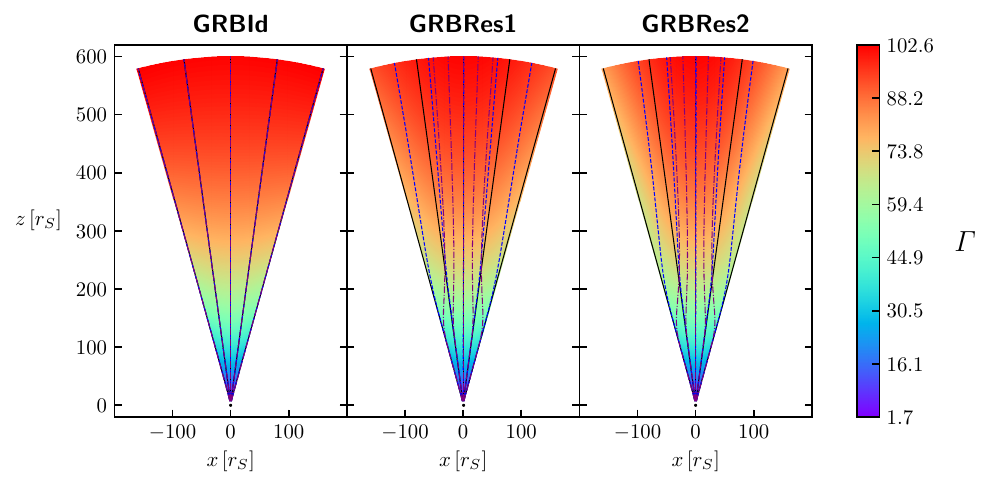}
		\caption{Spatial distribution of the Lorentz factor for the ideal MHD solution, GRBId and solutions GRBRes1 and GRBRes2. Black solid lines represent the magnetic flux lines, blue dashed lines correspond to contours of the total-energy-to-mass flux ratio, $\mu$, and purple dot-dashed lines are the contours of the electric potential, $\varPhi$. In the resistive solutions, the contours of $\varPhi$ start to diverge from the magnetic flux lines at the radial distance, $r_{\eta}$, from the black hole where the resistivity is switched on, and turn toward the jet axis. The contours of $\mu$ also diverge from the magnetic flux lines and turn toward the jet axis, although the modification of their geometry in the resistive solutions is not as significant as that of the contours of $\varPhi$.}  
		\label{lfmap}
	\end{figure*}

	In ideal MHD jets, contours of $\mu$ and of the electric potential, $\varPhi$, coincide with the magnetic flux lines on the poloidal plane. In the resistive jet solutions, the electric potential and, consequently, $\mu$ increase along the off-axis magnetic flux lines, and so the geometry of the aforementioned characteristic curves on the poloidal planes is altered. The deviation of the contours of $\varPhi$ and $\mu$, occurs over the dissipation regions, where Ohmic dissipation is significant. Inside these regions, both families of curves turn toward the jet axis, as both the electric potential and $\mu$ increase along the flux lines, shown in Fig. \ref{lfmap}. 
	
	The resistive solutions that we presented in this section suggest that resistivity on its own is not that significant in shaping the behavior of dissipative and resistive MHD outflows. The resistive solutions presented in this section all feature the same resistivity profile, given by Eq. \ref{res}. According to this profile, for $r > r_{\eta}$ the resistivity is nonzero in our solutions. Despite this, EM dissipation and deviation from ideal MHD in general is confined to a region of a specific length along the jet, instead of being present for all $r > r_{\eta}$. The reason for this is that dissipative effects emerge due to the electric field in the jet frame, $\bm{E}_{co}$, and are significant where $\bm{E}_{co}$ deviates significantly from zero. The electrical resistivity simply leads to the emergence of $\bm{E}_{co}$ through Ohm's law, by causing the variation of the poloidal magnetic field line angular velocity along the magnetic flux lines, which leads to the generation of an electric field component, or equivalently an electric potential difference, along the poloidal magnetic field lines. A nonzero resistivity everywhere in the solution does not necessarily mean that there will be a nonzero comoving electric field, $\bm{E}_{co}$, in the entirety of the solution. 
	
\begin{figure}
		\centering
		
		\includegraphics[width = 0.5\textwidth]{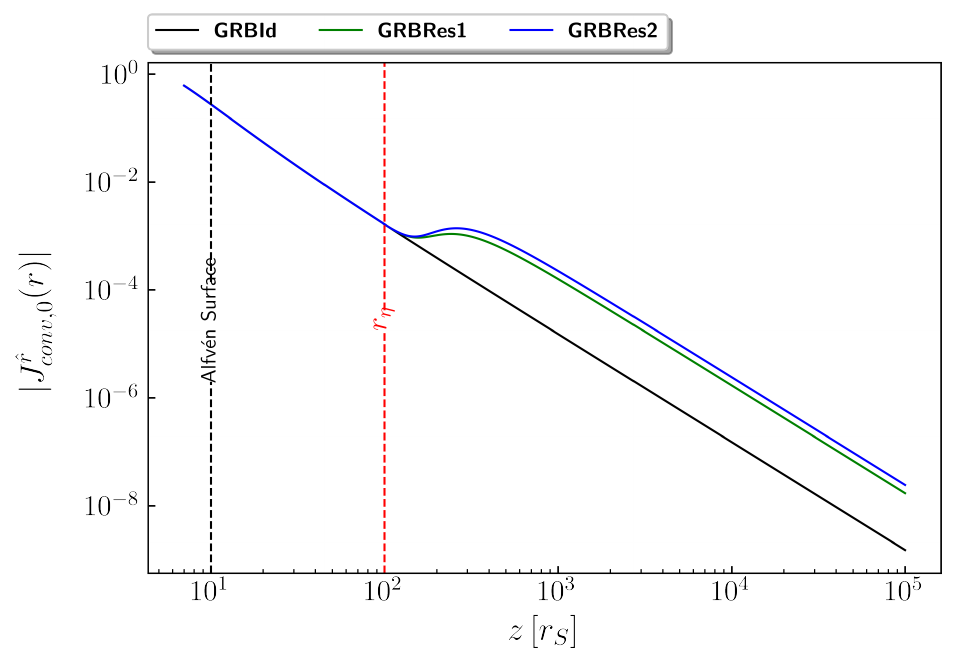}
		\caption{Convection current density along the jet axis. At $r_{\eta} = 100\, r_{S}$, the on-axis convection current shows an increase, which is a result of the increase of $\Omega$. This amplification of the radial convection current due to the increase of $\Omega$ over the dissipation regions is the cause behind the strengthening of the toroidal magnetic field.}  
		\label{jcond}
	\end{figure}

	This can be understood by considering the expression for the derivative of $\Omega$ along the poloidal field lines, Eq. \ref{dOmega}. The conduction current, $J_{cond,0}^{\hat{r}}(r)$ displays an increase over a short distance, before showing a steep decrease, following the behavior of the on-axis radial convection current $J_{conv,0}^{\hat{r}}(r)$, as portrayed in Fig. \ref{jcond}. 
	Consequently, the derivative of the field line angular velocity, $\Omega^{\prime}(r)$, also decreases rather rapidly with the radial distance, causing $\Omega^{\prime}(r)$ to drop to zero at a certain distance from $r_{\eta}$. As a result, the electric potential gradient along the poloidal magnetic field lines vanishes, and so does the comoving electric field, $\bm{E}_{co}$. For a certain resistivity then, a sufficient current density is required for the generation of considerable Ohmic dissipation in the jet.
	
	The decrease in the radial current density is the reason why placing $r_{\eta}$ at large distances from the jet base requires much larger resistivity values in order to have significant Ohmic dissipation, compared to resistivity profiles which are switched on at heights close to the jet base. As a general rule of thumb, we can say that the strength of Ohmic dissipation is determined by the product of the resistivity with the magnitude of the jet's radial current along its axis, $\eta_{2}(r)J_{0}^{\hat{r}}(r)$, as this is the quantity which determines the strength of the comoving electric field, $\bm{E}_{co}$. This is the reason why when we considered resistivity profiles which were contained over a particular region of the jet, such as a Gaussian profile, we obtained very similar results to the one we presented in this section. Of course a proper investigation of the significance of Ohmic dissipation requires taking into account both the specific enthalpy derivative and the Ohmic dissipation source term appearing in the entropy conservation equation (Eq. \ref{entropy}), as we did in the beginning of this section, where we defined the dimensionless number $R_{\beta}$.
	
	The stronger dissipation in the GRBRes2 resistive solution is due to the fact that this resistivity profile can attain higher values than the GRBRes1 solution over a shorter distance from $r = r_{\eta}$ where the resistivity is switched on in both solutions, due to its higher $\eta_{c}$ value. Solution GRBRes3, which features a resistivity profile characterized by a lower $\eta_{c}$ value than solutions GRBRes1 and GRBRes2, but displays a steeper increase with $r$, reproduces the behavior of the GRBRes2 solution of this section. The parameter $\sigma_{\eta}$ (Table \ref{tab:respar}) is the one that controls the resistivity profile's slope. A lower value means a steeper increase of the value of the resistivity with the distance from $r_{\eta}$.
	
\begin{figure}
		\centering
		
		\includegraphics[width = 0.5\textwidth]{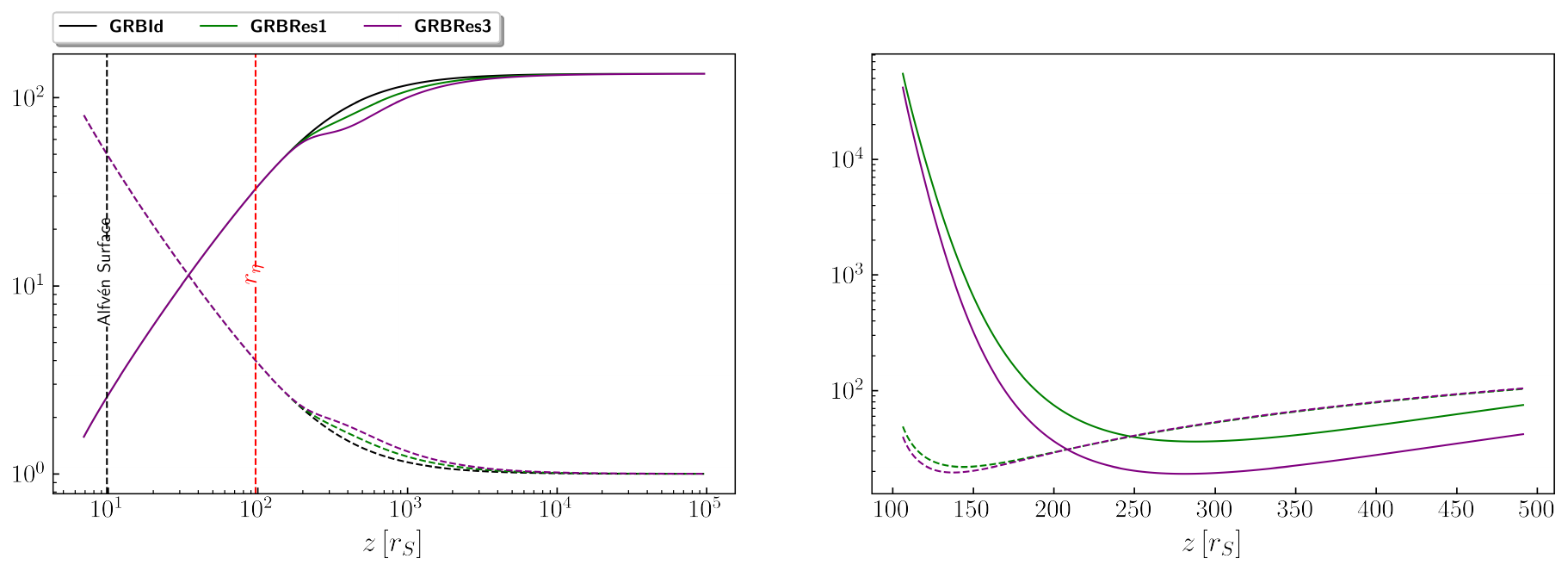}
		\caption{Comparison between solutions GRBRes1 and GRBRes3. Right panel: $-U_{t}$ (solid lines) and specific enthalpy (dashed lines) for solutions GRBId, GRBRes1, and GRBRes3. Left panel: $R_{\beta}$ (solid lines) and magnetic Reynolds number (dashed lines). Solution GRBRes3 presents a highly similar behavior to solution GRBRes2.}  
		\label{avlfgrb20}
	\end{figure}

	As can be seen in Fig. \ref{avlfgrb20}, $-U_{t}$ and the specific enthalpy of solution GRBRes3 is highly similar to that of solution GRBRes2. Additionally, solution GRBRes3 exhibits EM dissipation nearly as strong as the solution GRBRes2, as its resistivity profile increases more rapidly with the distance from $r_{\eta}$. These two resistivity profiles reach different constant values, although at a large enough distance from $r_{\eta}$ where dissipation has died out, due to the decrease in the on-axis radial convection current, as explained earlier. Over the dissipation region though, these two profiles have similar values, due to the more rapid increase of the GRBRes3 profile. The about equally strong dissipation exhibited by these two resistive solutions is illustrated by the similar $R_{\beta}$ profiles they display (Figs. \ref{avlfr} and \ref{avlfgrb20}). The rest of the nonideal effects we studied in this section also present strong similarities for these two resistive solutions.
	
	\subsection{Relative percent errors of the solutions}
	
\begin{figure*}
		\centering
		
		\includegraphics[width = 1.0\textwidth]{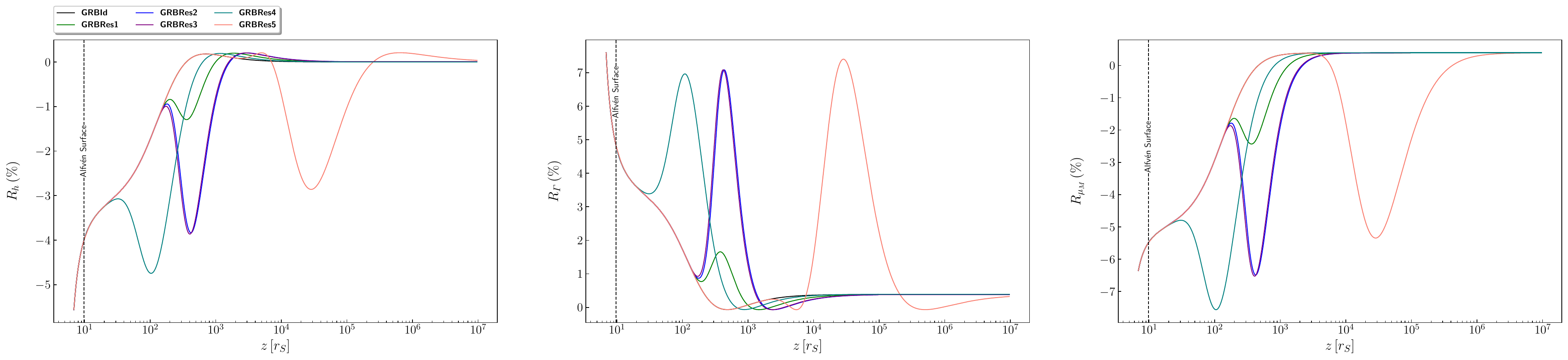}
		\caption{Relative percent errors in the angular expansions  for all the solutions presented in this section. Left panel: relative percent error in the angular expansion of the specific enthalpy, $R_{h}$. Middle panel: relative percent error in the angular expansion of the Lorentz factor, $R_{\varGamma}$. Right panel: relative percent error in the angular expansion of the matter component of the total energy flux-to-mass flux ratio, $R_{\mu_{M}}$.}  
		\label{R}
	\end{figure*}

	Finally, we calculated the relative percent errors in the angular expansions of the specific enthalpy, $R_{h}$, Lorentz factor, $R_{\varGamma}$, and matter component of the total energy flux-to-mass-flux ratio $R_{\mu_{M}}$, along the boundary streamline, for all five solutions presented in this section. The relative percent errors of these three quantities are displayed in Fig. \ref{R}. All five solutions satisfy the demand for validity, as the relative percent errors of all three quantities are below $10 \%$ along the jets' boundary streamlines.
	
	\section{Summary} \label{sec:5}
	
	In this work, we sought to understand the ways that nonzero electrical resistivity and Ohmic dissipation impacts the large scale properties and dynamics of relativistic astrophysical jets. In order to construct analytical models which describe resistive relativistic outflows of plasma in the vicinity of their axis of symmetry, we devised an analytical framework for the simplification of the partial differential equations which govern stationary and axisymmetric plasma outflows in the context of resistive relativistic MHD in Schwarzschild spacetime. The paraxial formalism we introduced allows the inclusion of a polytropic equation of state and of the relativistic generalization of Ohm's law in the system of the governing equations, while being able to reproduce the dynamics of ideal MHD relativistic jets. 
	
	The formalism we employed in the process of obtaining the solutions we presented is based on the assumption of suitable and physically motivated profiles for the fundamental quantities which describe relativistic jets. These profiles prescribe the dependency of these quantities on the polar angle, as measured from the jets' axis of symmetry. The equations of motion of the plasma alongside Maxwell's equations for the description of the jets' EM field, expressed in Schwarzschild spacetime, were expanded with respect to the polar angle, $\theta$. In the angular expansions of the governing equations, we retained terms of up to second order in $\theta$. The coefficients of the powers of $\theta$ which appear in the polynomials obtained from the angular expansion of the PDEs, make up the system of ODEs, the solution of which determines the dependence of all physical quantities on the radial distance from the central black hole. This system of ODEs has two singular points, which correspond to the sonic and Alfvén critical surfaces, which are spherical surfaces oriented along the radial direction, where the radial flow velocity is equal to the propagation velocity of sound and Alfvén waves in the jets respectively.
	
	Since our paraxial formalism is valid for polytropic outflows, we opted to accurately model the thermodynamics of the astrophysical outflows of interest, especially due to the Ohmic dissipation of EM energy which affects the thermodynamic properties of resistive jets. To that end, we chose to use the EoS introduced by \cite{ryu2006}, which, to our knowledge, is the most accurate analytical approximation of the correct Synge EoS for a single component relativistic gas \citep{synge1957}.
	
	To summarize, in the resistive MHD regime, the assumption of a nonzero resistivity necessitates the use of the relativistic generalization of Ohm's law, Eq. \ref{Ohm}. Thus, the "frozen-in" condition of ideal MHD breaks down and an electric potential gradient along the magnetic flux lines is generated. The angular expansion of the governing equations allowed us to determine this electric potential gradient along the streamlines as the source of dissipative phenomena in resistive jets. By considering the projection of Ohm's law along the flux lines, we discovered that the angular velocity of the rotation of the magnetic flux lines, $\Omega$, varies along them. The gradient of $\Omega$ along the flux lines is proportional to the product of the plasma's resistivity with the jet's Lorentz factor and the radial component of the conduction current. This gradient of $\Omega$ along the flux lines leads to the generation of the electric potential gradient along the jets' off-axis streamlines. Due to the electric potential gradient along the flux lines, an electric field component parallel to the poloidal magnetic field forms. This parallel component to the poloidal magnetic field alters the geometry of the electric field from the one predicted by the "frozen-in" condition of ideal MHD. Thus, a comoving electric field is generated, which leads to Ohmic dissipation. Ohmic dissipation does not affect the jet axis directly, but increases away from the axis as $\theta^{2}$, and its strength is determined by the electric potential gradient along the magnetic flux lines.
		
	The ideal MHD solutions we obtained are thermally accelerated, due to the conversion of their thermal internal energy to kinetic energy of the plasma. At short distances from their bases, where $\Theta \gg 1$, the strongly relativistic solutions, AGNId and GRBId accelerate as $\varGamma\sim z$ ($z \simeq r$ for small $\theta$), with our solutions reproducing the acceleration profile predicted by the thermally accelerated relativistic fireball model \citep{shemi1990}. At larger distances, where the plasma's thermal energy has decreased significantly, the acceleration profile of all three solutions becomes smoother, before the thermal acceleration ends and the jets obtain their asymptotic terminal Lorentz factors. 
		
	The shapes and collimation profiles of the jets are determined by the transfield components of the thermal pressure gradient and of the Lorentz force. The strongly relativistic solutions experience weaker to negligible collimation, compared to the weakly relativistic solution, XRBId. Asymptotically, the streamlines of the strongly relativistic solutions, AGNId and GRBId, expand freely and their poloidal magnetic fields obtain a radial configuration. The negligible collimation of the strongly relativistic solutions is a consequence of the decollimating transfield electric force, which effectively cancels out the collimating effect of the magnetic pinching force of the toroidal magnetic field. 
		
	Non-zero electrical resistivity and the associated Ohmic dissipation mainly impact relativistic jets in three areas: acceleration, collimation, and field topology. Firstly, Ohmic dissipation converts the energy stored in the EM field to thermal energy. This reduces the rate at which the plasma's thermal internal energy is converted to kinetic energy, while in cases of particularly strong dissipation, even deceleration of the jet is possible, as in the case of solution GRBRes5. In this particular solution, the resistivity is switched on at the end of the thermal acceleration region, where most of the jet's thermal energy has been converted to kinetic energy and as such the jet is cold. The thermal acceleration mechanism is more strongly impacted the further away we move from the jet's axis, as the dissipated power per unit volume is proportional to the square of the polar angle. Thus, the thermal acceleration mechanism is more strongly affected near the jet boundary. The thermal acceleration of the plasma along the jet axis is not directly affected by Ohmic dissipation, as the latter is not present there.
		
	Resistive jets experience stronger collimation than their ideal MHD counterparts. Due to Ohmic dissipation, streamlines away from the axis experience stronger dissipative heating than streamlines with smaller opening angles. This intensifies the collimating transfield thermal pressure gradient. Additionally, the transfield component of the collimating magnetic force is stronger in resistive jets, due to the amplification of their magnetic field and radial current density. However, the differences in the collimation profiles and shapes of the poloidal field lines between the resistive solutions and their ideal MHD counterparts are not significant. Nevertheless, the slightly stronger collimation that resistive solutions experience is able to indirectly affect their acceleration. Their streamlines do not expand as freely as in ideal MHD jets characterized by the same parameters, reducing the rate at which their thermal energy is converted to kinetic. While this effect is practically negligible, it affects the on-axis acceleration of the resistive jets, even though no Ohmic dissipation emerges along their axes.
		
	The toroidal magnetic field is significantly amplified in resistive jets. This alters the geometrical configuration of their 3D magnetic field considerably, with the magnetic field not following the classical scaling $|B^{\hat{\phi}}|\sim z^{-1}$ predicted by ideal MHD. The reason for the amplification of the toroidal magnetic field in resistive jets, is once again the increase of $\Omega$, or equivalently the electric potential increase, along the  lines of the poloidal magnetic field. The increase in $\Omega$ over the dissipation region in resistive jets leads to an increase in the charge density. Due to this increase in the charge density, the radial component of the convection current density, which dominates over the conduction current, is amplified, and thus, the toroidal magnetic field is strengthened. The poloidal magnetic field is not affected in any significant way.
		
	The electric potential gradient which appears along the magnetic flux lines alters the shape of the electric potential contours on the poloidal plane, so that they do not coincide with the lines of poloidal magnetic field and velocity, as in ideal MHD jets. The electromagnetic component of the total energy flux-to-mass flux ratio, $\mu_{EM}$ is similarly affected. This alters the shape of the contours of $\mu$ on the poloidal plane, too, causing them to turn toward the inner parts of the jet over the length of the region where Ohmic dissipation is significant. As dissipation wanes with the distance from the jet base, the electric potential contours and $\mu$ contours become parallel to the streamlines.
		
	In the resistive solutions we presented, the resistivity follows a $\tanh$ profile with the radial distance, initialized from zero at a radial distance $r_{\eta}$ from the central object. Even though in all resistive solutions the resistivity reaches a constant, nonzero value $\eta_{c}$, Ohmic dissipation is significant only over a certain distance along the jet from the position where the resistivity is switched on. These dissipation regions are the regions where the electric potential gradient along the flux lines is present. The reason for their finite length is the dependence of the field line angular velocity gradient along the flux lines on the radial component of the conduction current. This current decreases with distance, due to the weakening of the jets' electromagnetic field the further away we move from the jet base, decreasing the gradient of $\Omega$ along the field lines. As the gradient of $\Omega$ along the flux lines decreases and asymptotically becomes zero, dissipative effects weaken and consequently die out after a certain distance from $r_{\eta}$.
		
	As the dissipation regions of resistive relativistic jets are characterized by an electric potential gradient along the poloidal magnetic field lines, they can act as extended sites where the acceleration of charged particles due to the parallel electric field can take place. The significance and efficiency of this particle acceleration mechanism needs to be the subject of future investigations.

	\begin{acknowledgements}
		We wish to thank the anonymous referee for providing valuable feedback which helped improve the quality of the paper. This work was supported in full by the State Scholarships
		Foundation (IKY) scholarship program from the proceeds of the “Nic. D.
		Chrysovergis” bequest.
	\end{acknowledgements}
	
	%
	%
	
	\bibliographystyle{aa}
	\bibliography{references}
	
	\onecolumn
	\begin{appendix}
		\section{Angular expansions}\label{sec:angular}
		In this Appendix, we present the expressions obtained from the angular expansions of fundamental jet scalar quantities and vectors fields. 
		
		The expansion of the magnetic field, Eq. \ref{mag}, with respect to $\theta$ can be expressed in terms of the expansion factor, $F(r)$, as
		\begin{equation}
			\tilde{\bm{B}}(r,\theta) = \dfrac{A_{2}(r)}{r^{2}}\left(2 - \theta^{2}\right)\bm{\hat{r}} - \dfrac{h_{t}(r)A_{2}(r)F(r)}{r^{2}}\theta\bm{\hat{\theta}} + B^{\hat{\phi}}_{1}(r)\theta\bm{\hat{\phi}}\, ,
		\end{equation}
		while the expansion of $\bm{u}$, Eq. \ref{vel}, is
		\begin{equation}
			\tilde{\bm{u}}(r,\theta) = \dfrac{\Psi_{A}A_{2}(r)}{4\pi\rho_{0}(r)r^{2}}\left(2 - \left(1 + \dfrac{2\rho_{2}(r)}{\rho_{0}(r)}\right)\theta^{2}\right)\bm{\hat{r}} - \dfrac{\Psi_{A}h_{t}(r)A_{2}(r)F(r)}{4\pi\rho_{0}(r)r^{2}}\theta\bm{\hat{\theta}} +  u^{\hat{\phi}}_{1}(r)\theta\bm{\hat{\phi}}\, .
		\end{equation}
		The expansion of the electric field, Eq. \ref{el}, using Eq. \ref{phi2}, is written as
		\begin{equation}\label{efield}
			\tilde{\bm{E}}(r,\theta) = -\left(\dfrac{\Omega(r)F(r)A_{2}(r)}{c r} + \dfrac{\Omega^{\prime}(r)A_{2}(r)}{c}\right)\theta^{2}\bm{\hat{r}} - \dfrac{2A_{2}(r)\Omega(r)}{ch_{t}(r)r}\theta\bm{\hat{\theta}}\, .
		\end{equation}
		The expansion of the charge density, $J^{0}/c$ is
		\begin{equation}
			\tilde{J}^{0}(r,\theta) = J^{0}_{0}(r) + J^{0}_{2}(r)\theta^{2}\, ,
		\end{equation}
		where
		\begin{equation}\label{charge}
			J^{0}_{0}(r) = -\dfrac{\Omega(r)A_{2}(r)}{\pi r^{2}h_{t}(r)}\, ,
		\end{equation}
		\begin{equation}
			J^{0}_{2}(r) = \dfrac{J^{0}_{0}(r)}{4}\left(h_{t}(r)^{2}F(r)\left(F(r) + \derivative{\ln{F(r)}}{\ln{r}} + 1\right) + h_{t}(r)^{2}\derivative{\ln{\Omega(r)}}{\ln{r}}\left(2F(r) + \derivative{\ln{\Omega^{\prime}(r)}}{\ln{r}} + 2\right) - 6\right)\, ,
		\end{equation}
		The angular expansion of the current density has the form
		\begin{equation}
			\tilde{\bm{J}}(r,\theta) = \left(J_{0}^{\hat{r}}(r) + J_{2}^{\hat{r}}(r)\theta^{2}\right)\bm{\hat{r}} + J_{1}^{\hat{\theta}}(r)\theta\bm{\hat{\theta}} + J_{1}^{\hat{\phi}}(r)\theta\bm{\hat{\phi}}\, ,
		\end{equation}
		with
		\begin{equation}\label{current}
			\begin{split}
				J_{0}^{\hat{r}}(r)& = \dfrac{c B_{1}^{\hat{\phi}}(r)}{2\pi r},\,\, J_{2}^{\hat{r}}(r) = -\dfrac{c B_{1}^{\hat{\phi}}(r)}{4\pi r},\,\, J_{1}^{\hat{\theta}}(r) = -\dfrac{c h_{t}(r)B^{\hat{\phi}}_{1}(r)}{4\pi r}\left(\derivative{\ln{B^{\hat{\phi}}_{1}(r)}}{\ln{r}} + \derivative{\ln{h_{t}(r)}}{\ln{r}} + 1\right),\\\\
				J_{1}^{\hat{\phi}}(r)& =-\dfrac{c A_{2}(r)}{4\pi r^{3}}\left(h_{t}(r)^{2}F(r)\left(F(r) + \derivative{\ln{F(r)}}{\ln{r}} + 2\derivative{\ln{h_{t}(r)}}{\ln{r}} - 1\right) - 2\right)
			\end{split}
		\end{equation}
		
		The angular expansion of the Lorentz factor is
		\begin{equation}
			\tilde{\varGamma}(r,\theta) = \varGamma_{0}(r) + \varGamma_{2}(r)\theta^{2}\, ,
		\end{equation}
		with
		\begin{equation}
			\varGamma_{0}(r) = \dfrac{\sqrt{\Psi_{A}^{2}A_{2}(r)^{2} + 4\pi^{2}c^{2}r^{4}h_{t}(r)^{2}\rho_{0}(r)^{2}}}{2\pi c r^{2}h_{t}(r)\rho_{0}(r)}\, ,
		\end{equation}
		\begin{equation}
			\varGamma_{2}(r) = \dfrac{16\pi^{2}r^{4}u^{\hat{\phi}}_{1}(r)^{2}\rho_{1}(r)^{3} + \Psi_{A}^{2}A_{2}(r)^{2}\left((h_{t}(r)^{2}F(r)^{2} - 4)\rho_{0}(r) - 8\rho_{2}(r)\right)}{32\pi^{2}c^{2}r^{4}h_{t}(r)^{2}\varGamma_{0}(r)\rho_{1}(r)^{3}}\, .
		\end{equation}
		The dimensionless temperature is
		\begin{equation}
			\tilde{\Theta}(r,\theta) = \Theta_{0}(r) + \Theta_{2}(r)\theta^{2}\, ,
		\end{equation}
		where
		\begin{equation}
			\Theta_{0}(r) = \dfrac{P_{0}(r)}{\rho_{0}(r)c^{2}}\, ,
		\end{equation}
		\begin{equation}
			\Theta_{2}(r) = \dfrac{P_{2}(r)}{\rho_{0}(r)c^{2}} - \dfrac{P_{0}(r)\rho_{2}(r)}{\rho_{0}^{2}c^{2}}
		\end{equation}
		The expression obtained from the angular expansion of the specific enthalpy is
		\begin{equation}
			\tilde{\xi}(r,\theta) = \xi_{0}(r) + \xi_{2}(r)\theta^{2}\, ,
		\end{equation}
		where the functions, $\xi_{0}(r),\, \xi_{2}(r)$, are
		\begin{equation}
			\xi_{0}(r) = \dfrac{12P_{0}(r)^{2} + 8P_{0}(r)\rho_{0}(r)c^{2} + 2\rho_{0}(r)^{2}c^{4}}{3P_{0}(r)\rho_{0}(r)c^{2} + 2\rho_{0}(r)^{2}c^{4}}\, ,
		\end{equation}
		\begin{equation}
			\xi_{2}(r) = \dfrac{36P_{0}(r)^{2} + 48P_{0}(r)\rho_{0}(r)c^{2} + 10\rho_{0}(r)^{2}c^{4}}{(3P_{0}(r) + 2\rho_{0}(r)c^{2})^{2}}\Theta_{2}(r)\, .
		\end{equation}
		
		The angular expansions of the unit vectors along the streamlines, $\bm{\hat{t}}$, and in the transfield direction, $\bm{\hat{n}}$, are
		\begin{equation}
			\bm{\hat{t}} = \left(1 - \dfrac{h_{t}(r)^{2}F(r)^{2}}{8}\theta^{2}\right)\bm{\hat{r}} - \dfrac{h_{t}(r)F(r)}{2}\theta\bm{\hat{\theta}}
		\end{equation}
		\begin{equation}
			\bm{\hat{n}} = \dfrac{h_{t}(r)F(r)}{2}\theta\bm{\hat{r}} + \left(1 - \dfrac{h_{t}(r)^{2}F(r)^{2}}{8}\theta^{2}\right)\bm{\hat{\theta}} 
		\end{equation}
		The angular expansions of the thermal pressure gradient, $\bm{f}_{P}$, and of the Lorentz force, $\bm{f}_{L}$, in the transfield direction are
		\begin{equation}
			f_{P\perp}(r,\theta) = f_{P1}(r)\theta\, ,
		\end{equation}
		and
		\begin{equation}
			f_{L\perp}(r,\theta) = f_{L1}(r)\theta\, .
		\end{equation}
		The angular expansions of the force due to the specific enthalpy gradient ("temperature" force), $\bm{f}_{T}$ and of the component of the Lorentz force along the streamlines have the forms
		\begin{equation}
			f_{T}(r,\theta) = f_{T0}(r) + f_{T2}(r)\theta^{2}\, ,
		\end{equation}
		and 
		\begin{equation}
			f_{L\parallel}(r,\theta) = f_{L2}(r)\theta^{2}\, .
		\end{equation}
		The expressions of $f_{P1}(r)$, $f_{L1}(r)$, $f_{T0}(r)$, $f_{T2}(r)$, and $f_{L2}(r)$ are not as fundamental as the polynomial coefficients in the previous quantities. Additionally, some of these expressions are excessively long, which is why we chose not to include them in the text.
	\end{appendix}
	
\end{document}